\documentclass[12pt]{article}
\usepackage[utf8]{inputenc}
\usepackage{amsthm,amsmath,amsfonts,amssymb,bm,physics}
\usepackage{graphicx}
\usepackage{apacite}
\usepackage{float}
\usepackage{subfigure}
\usepackage{rotating} 
\usepackage{algorithm}
\usepackage{algorithmic}
\usepackage{natbib}
\usepackage{url}
\usepackage[margin=1in]{geometry}
\usepackage{setspace}
\usepackage{tabularx}
\usepackage{booktabs}
\usepackage{authblk}
\usepackage{indentfirst}
\usepackage{xcolor}
\usepackage{times}
\newtheorem{theorem}{Theorem}[section]

\theoremstyle{remark}
\renewcommand{\bar}{\overline}
\renewcommand{\hat}{\widehat}

\usepackage{xr}
\usepackage{xr-hyper}
\usepackage[colorlinks,citecolor=blue,urlcolor=blue]{hyperref}
\usepackage{comment}
\usepackage{enumerate}
\usepackage{enumitem}
\usepackage{setspace}
\usepackage[normalem]{ulem}
\def\spacingset#1{\renewcommand{\baselinestretch}%
{#1}\small\normalsize} \spacingset{1.5}
\def\blue{\color{blue}}
\def\red{\color{red}}
\def\mR{\mathbb{R}}
\def\M{\bm{M}}
\def\Q{\bm{Q}}
\def\X{\bm{X}}
\def\defeq{\stackrel{\mathrm{def}}{=}}  % for definitions
\newcommand{\bfsym}[1]{\ensuremath{\boldsymbol{#1}}}
\def\balpha{\bfsym \alpha}
\def\bbeta{\bfsym \beta}
\def\bgamma{\bfsym \gamma}             \def\bGamma{\bfsym \Gamma}
\def\bdelta{\bfsym {\delta}}           \def\bDelta {\bfsym {\Delta}}
\def\bfeta{\bfsym {\eta}}              \def\bfEta {\bfsym {\Eta}}
\def\bmu{\bfsym {\mu}}                 \def\bMu {\bfsym {\Mu}}
\def\bnu{\bfsym {\nu}}
\def\btheta{\bfsym {\theta}}           \def\bTheta {\bfsym {\Theta}}
\def\beps{\bfsym \varepsilon}          \def\bepsilon{\bfsym \varepsilon}
\def\bsigma{\bfsym \sigma}             \def\bSigma{\bfsym \Sigma}
\def\blambda {\bfsym {\lambda}}        \def\bLambda {\bfsym {\Lambda}}
\def\bomega {\bfsym {\omega}}          \def\bOmega {\bfsym {\Omega}}
\def\brho   {\bfsym {\rho}}
\def\btau{\bfsym {\tau}}
\def\bxi{\bfsym {\xi}}          \def\bXi{\bfsym {\Xi}}
\def\bzeta{\bfsym {\zeta}}
\def\mY{\mathbb{Y}}
\def\mZ{\mathbb{Z}}
\def\mX{\mathbb{X}}
\def\mR{\mathbb{R}}
\def\bX{\mathbf{X}}
\def\zero{\mathbf{0}}
\def\mG{\mathcal{G}}

\def\red{\color{red}}
\def\blue{\color{blue}}

\def\1{{\bf 1}}

\renewcommand{\bbeta}{\mbox{\boldmath $\beta$}}
\renewcommand{\bdelta}{\mbox{\boldmath $\delta$}}

\begin{document}

%\input{TitlePage}
%\newpage

\title{\Large\bfseries Linking COPD Prevalence with Income Distribution:
A Spatial Heterogeneous Compositional Regression via Geographically Weighted Penalized Approach}

%\author{Jingwen Deng, Shujie Ma, Sergio J. Rey, Guanyu Hu}
\author{
Jingwen Deng$^1$\thanks{ORCID: 0009-0007-6673-1892}, Shujie Ma$^2$, Sergio J. Rey$^3$, Guanyu Hu$^4$
}
\date{}
\maketitle
\begin{center}
{\small
$^1$ Department of Biostatistics and Data Science, The University of Texas Health Science Center at Houston\\
$^2$ Department of Statistics, University of California, Riverside\\
$^3$ Department of Geography, San Diego State University\\
$^4$ Department of Statistics and Probability, Michigan State University
}
\end{center}
%\author{}
%\date{\today}
\date{}
\maketitle

%\begin{abstract}
\iffalse
    Income inequality, a key driver of health disparities, is often measured compositionally (e.g., income distribution), with its effect varying spatially. Traditional spatial models fail to capture abrupt spatial heterogeneity, suffer from computational limits in high dimensions, and inadequately address the unique constraints of compositional covariates. To address these issues, we propose a novel geographically weighted penalized regression model that incorporates compositional covariates effectively. Our approach utilizes a geographically weighted pairwise fusion penalty to identify both contiguous and discontiguous clusters in regression coefficients across regions, relaxing the restrictive assumption of spatial smoothness or strict contiguity. By employing a penalty function, such as MCP, our framework improves estimation accuracy and model interpretability in complex spatial settings. We demonstrate the method's utility through an empirical analysis of Income Inequality and COPD prevalence. This innovation provides a robust, scalable tool for spatial data analysis with compositional covariates, enabling more accurate inference on geographically varying phenomena
    \fi
    Income inequality is a major contributor to health disparities, yet its effects often vary by geography and are commonly represented as compositional distributions (e.g., proportions of households across income brackets). Existing spatial regression methods struggle in this setting: they typically assume smooth spatial variation, cannot accommodate abrupt spatial heterogeneity, and lack principled treatment of compositional covariates. We propose a geographically weighted penalized compositional regression model that addresses these challenges simultaneously. Our method adopts a pairwise fusion penalty that enables detection of both contiguous and noncontiguous regional clusters with shared regression effects, thereby relaxing strong assumptions of spatial smoothness {and geographic contiguity}. This allows regions with similar underlying socioeconomic structures to be identified even when they are not geographically adjacent. By incorporating nonconvex penalties, such as the {minimax concave penalty (MCP)}, the approach achieves improved estimation accuracy, interpretability, and scalability in high-dimensional spatial settings. We illustrate the method through an analysis linking U.S. income composition to {chronic obstructive pulmonary disease (COPD)} prevalence, revealing spatially heterogeneous associations that are obscured by conventional models. The proposed framework provides a flexible and robust tool for spatial data analysis involving compositional predictors and region-specific heterogeneity.
%\end{abstract}

\noindent
{Keywords:} Health Disparities, Penalized Regression, Spatial Clustering, Socioeconomic Factors

% \begin{itemize}
%    \item Structure of introduction: 1. introduce compostional covariate and COPD response. 2. introduce exiting  spatially varying models and spatial clustered model. 3. introduce chanllanging in exising methods. 4. introduce our proposed model and advantages. 5. summary the structure of our paper.
%     \item Motivatitng data: get table 1 for our data and visualize our data in one figure. 
%     \item Method section: 1. introduce compostitional regression model with log-contrast regression. 2. introduce our transformation. 3. introduce our geographically weighted panalty and discuss the selections of spatial weights. 4. introduce our full panalized regression model 5. our admm algorithm and BIC selection of panlety
%     \item Simulation: 1. introduce simulation settings with figures (spatial partitions) table (parameters for each design) 2. evaluation metric (clustering metric like MCR and Rand index and estimation performance metric). 3. present simulation results
%     \item real data analysis: introduce the model and covariates for our real data analysis: discuss the population adjustment and present estimation results and interpret results based on estimated coefficients and clusters. In addition, you should check some common characteristics of the counties or states belong to same clusters and have one paragraph to discuss that. You could let ChatGPT tell you some shared features of the counties or the states in the same clusters.
%     \item disucssion section: summary our findings and constributions and list two or three future works. 
% \end{itemize}
\section{Introduction}

%\subsection{Overview} 
\label{subsec:Overview}

{Income inequality is a well-recognized driver of health disparities, with its effects often varying across geographic regions \citep{Subramanian_2004_ineq}. These disparities are often represented through compositional measures such as income distribution, capturing the relative structure of socioeconomic groups within a region. 
A key challenge in this context is that the relationship between socioeconomic factors and health outcomes is inherently spatially heterogeneous. While Tobler’s First Law of Geography \citep{Tobler_1970_computer} suggests that nearby regions tend to be more similar, many real-world processes exhibit more complex spatial structures, including abrupt changes and similarities across non-adjacent regions. For example, regions with comparable socioeconomic conditions or health system characteristics may display similar patterns in disease prevalence even when they are geographically distant.}

{Chronic obstructive pulmonary disease (COPD), a major public health concern characterized by high prevalence and substantial healthcare burden, provides an important case for studying such spatial heterogeneity. Although prior research \citep{quin_2020_income} has linked income inequality to health outcomes, the role of income composition in shaping spatial variation in COPD prevalence remains insufficiently understood.
These challenges highlight the need for analytical frameworks that can capture both compositional dependencies and complex spatial structures, enabling a more nuanced understanding of how socioeconomic factors contribute to geographic disparities in health outcomes.}

{Existing approaches to analyzing spatial data and their impact on health outcomes can be broadly categorized into two main types: spatially varying coefficient models (SVCMs) and spatially clustered coefficient models (SCCMs).\footnote{For a recent overview see \cite{anselin2024EndogenousSpatial}.}
SVCMs, including geographically weighted regression \citep[GWR,][]{brunsdon_geographically_1996}, Gaussian process-based models \citep{gelfand_spatial_2003}, and related spatially varying coefficient frameworks, capture spatial heterogeneity through smoothly varying coefficients, making them well-suited for settings with gradual spatial transitions. 
In contrast, SCCMs aim to identify regions with homogeneous regression relationships. These include classical clustering methods (e.g., K-means and hierarchical clustering \citep{macqueen_methods_1967, giraldo_hierarchical_2012}), model-based approaches (e.g.,Gaussian mixture models \citep[GMMs,][]{everitt_finite_1981, banfield_model-based_1993, tibshirani_regression_1996, mcnicholas_model-based_2010, wei_latent_2013}, logistic-normal mixture models \citep{shen_inference_2015}), and Bayesian nonparametric methods (e.g., Dirichlet process mixture models \citep[DPMMs,][]{suarez_bayesian_2016}), along with spatially constrained variants, such as Markov random fields \citep[MRFs,][]{jiang_clustering_2012} and Gaussian Markov random fields \citep[GMRFs,][]{prates_transformed_2015}}

{Despite their contributions, existing approaches face several important limitations. First, SVCMs rely on smoothness assumptions that may fail to capture abrupt spatial changes or clearly defined regional boundaries, which are common in many geographic processes. This can lead to biased or overly smoothed representations of spatial heterogeneity. 
Second, as the number of covariates or spatial units increases, these models can become computationally intensive and prone to overparameterization, limiting their applicability in large-scale geographic studies. 
Third, many spatial clustering approaches \citep{lee_cluster_2017,li2019spatial,kamenetsky2022regularized,wang2024scanner} enforce strict geographic contiguity, which can obscure meaningful similarities across non-adjacent regions. In practice, regions that share similar socioeconomic or health-related characteristics are not necessarily geographically contiguous, making contiguity-based clustering restrictive for understanding broader spatial patterns.
Fourth, handling compositional covariates (e.g., income distributions) introduces additional challenges due to their inherent dependence structure under the constant-sum constraint \citep{greenacre2021compositional}. Standard spatial models often fail to account for this structure, potentially leading to invalid inference or misleading interpretations.
Finally, many clustering-based methods require pre-specifying the number of clusters or rely on strong distributional assumptions, which can limit flexibility and affect interpretability in complex spatial settings.
These limitations are particularly restrictive from a geographic perspective, where regions with similar socioeconomic or health-related structures may not be geographically contiguous but still share common underlying processes. %Capturing such non-contiguous yet structurally similar regions is essential for understanding spatial inequality and regional heterogeneity beyond proximity-based assumptions.
}

{To overcome these limitations, we propose a novel Geographically Weighted Penalized Compositional Regression model that bridges spatially varying and spatial clustering frameworks. The proposed approach introduces a geographically weighted pairwise fusion penalty that enables the identification of both locally contiguous and spatially discontinuous clusters of regression coefficients, including those across non-adjacent regions. By relaxing strict geographic contiguity assumptions, the model allows spatial regimes to be defined based on structural similarity rather than proximity. This is particularly important in geographic analysis.}

{the framework further accommodates compositional predictors by respecting their sum-to-one constraint and simplex geometry through log-ratio transformations, ensuring statistically valid inference when modeling proportional data such as income distributions. The incorporation of non-convex penalties, such as the minimax concave penalty \citep[MCP,][]{zhang_nearly_2010}, further improves estimation accuracy while maintaining interpretability in high-dimensional settings. }

{Taken together, these features provide a flexible and data-driven framework for uncovering spatial heterogeneity and regional inequality. The proposed method enables the identification of spatial patterns that reflect underlying relationships among regions, offering new insights into how compositional socioeconomic factors are associated with variation in health outcomes.}

The remainder of this paper is organized as follows. Section \ref{sec:motivation_data} introduces the motivating data. Section \ref{sec:method} presents the proposed methodology, including the model framework and estimation procedure. Section \ref{sec:simulation} evaluates the performance of the proposed approach through simulation studies. Section \ref{sec:sensitivity} examines the sensitivity of the clustering results to the decay parameter $r$. Section \ref{sec:real_data} provides empirical analyses. Finally, Section \ref{sec:discussion} concludes with a discussion of implications, limitations, and future directions. Technical details, including the Alternating Direction Method of Multipliers algorithm \citep[ADMM,][]{boyd_distributed_2010} and hyperparameter selection, as well as additional simulation results, are provided in the Appendix.

\section{Motivating data}\label{sec:motivation_data}

% start here
This study explores the spatial variability in COPD prevalence on two distinct geographic scales. It includes a state-level dataset that spans all 51 US states, offering insights into socioeconomic and health disparities at a national level. In addition, a county-level dataset examines 254 counties within Texas, providing detailed local analysis. This dual-scale approach allows for the examination of potential clustering or explanatory patterns that may become evident when data is viewed at broader state levels versus more detailed county levels.

\begin{figure}[tbp]
    \centering
    \includegraphics[width=5.5in]{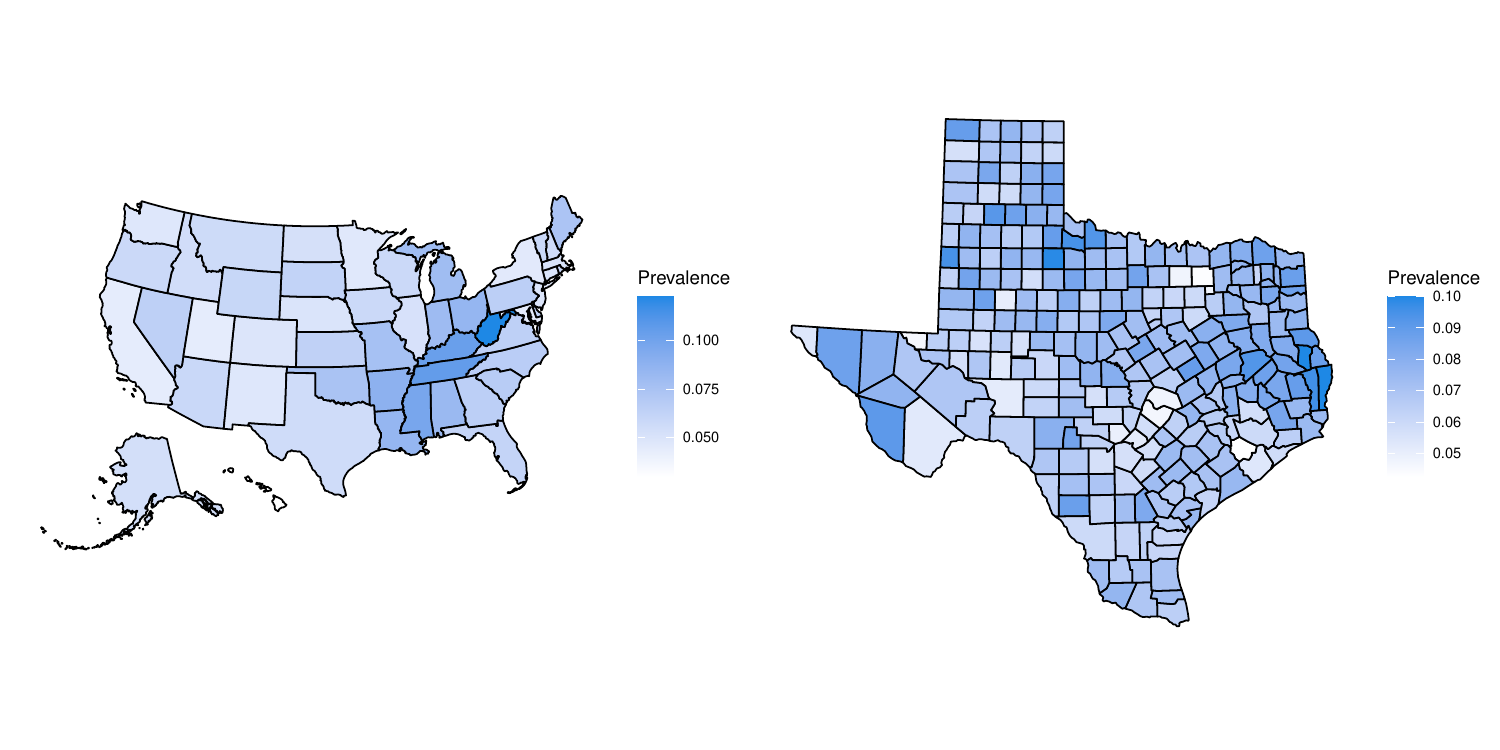}
    \caption{Estimated COPD prevalence across the United States by state (left) and within Texas by county (right). The color gradient reflects the percentage of adults with COPD in each region.}
    \label{fig:COPD2}
\end{figure}

The prevalence data for COPD were obtained from the Centers for Disease Control and Prevention's Local Data for Better Health project~\citep{cdc_places_2024}. The state data includes age-adjusted COPD rates for the year 2022. In contrast, the county dataset for Texas incorporates model-based estimates derived from the 2021 Behavioral Risk Factor Surveillance System~\citep{cdc_brfss_2024}, data from the 2017–2021 American Community Survey (ACS), and 2021 population estimates from the Census. Figure~\ref{fig:COPD2} illustrates the geographic distribution of these estimates, highlighting regional differences both across the United States (left panel) and within Texas (right panel).

A critical aspect of this analysis is the investigation of compositional data concerning the annual household income distribution for 2022, which was obtained from \citet{us_census_acs_2024}. This data is segmented into four income brackets, ranging from low to high, based on standardized definitions. Given that the proportions total one hundred percent, the interdependence inherent in compositional data necessitates specific analytical techniques.

Furthermore, Table~\ref{tab:table1_ustx} presents key descriptive statistics for non-compositional covariates across each geographic scale. At the state level, this includes: (a) GDP in manufacturing; (b) obesity rate, defined as the proportion of adults with a body mass index (BMI) of 30 or higher; (c) arthritis prevalence; and (d) smoking rate, which is the proportion of adults who currently smoke. Data on obesity rates, arthritis prevalence, and smoking rates were sourced from \citet{cdc_places_2024}, and GDP in manufacturing data came from \citet{bea_gdp_state_2024}. The county-level dataset includes: (a) the percentage of Asian population; (b) the percentage of the population residing in rural areas; (c) the percentage of children under age 18; and (d) the percentage of children eligible for free or reduced price lunch; (e) obesity rate and (f) county-specific premature age-adjusted mortality rates, indicative of overall health outcomes. All county-level data were retrieved from \citet{chr_2022}.

\renewcommand{\arraystretch}{1}
\setlength{\tabcolsep}{6pt}
\small
\begin{table}[tbp]
\centering
\caption{\label{tab:table1_ustx} Descriptive statistics for variables included in the state-level and county-level study, categorized into Socioeconomic Factors and Health Factors. Mean (SD) and Median [Min, Max] values are shown for each variable.} 
\vspace{0.1cm}
\resizebox{5in}{!}{
\begin{tabular}{p{5cm}|r|r}
\toprule
\textbf{Variable} & \textbf{State Level (n = 51)} & \textbf{County Level (n = 254)} \\    
\midrule
\multicolumn{3}{l}{\textbf{Outcome Variable}} \\    
\midrule
COPD prevalence (\%) &  &  \\ 
\hspace{5mm} Mean (SD) & 6.39 (1.81) & 7.16 (1.08) \\ 
\hspace{5mm} Median [Min, Max] & 6.00 [3.00, 12.30] & 7.15 [4.30, 10.00] \\ 
\midrule
\multicolumn{3}{l}{\textbf{Socioeconomic Factors}} \\   
\midrule
Income proportion (\%) & & \\         
\hspace{5mm}Low & 22.30 (5.26) & 21.60 (7.13) \\
\hspace{5mm}Low-medium & 17.80 (2.92) & 23.70 (5.67) \\
\hspace{5mm}Medium & 31.10 (3.13) & 29.90 (4.91) \\
\hspace{5mm}High & 28.80 (7.46) & 24.80 (8.75) \\
Manufacturing GDP & 52000 (68300) & - \\ 
Asian (\%) & - & 1.40 (2.14)  \\
Rural (\%) & - & 55.52 (31.90)  \\
Below 18 (\%) & - & 23.64 (3.86) \\
Children free lunch (\%) & - & 60.65 (13.51) \\
\midrule
\multicolumn{3}{l}{\textbf{Health Factors}} \\    
\midrule
Adult Obesity (\%) & 33.80 (4.06) & 37.81(2.90) \\
Arthritis (\%) & 27.80 (3.92) & - \\ 
Smoking (\%) & 13.70 (2.83) & - \\ 
Premature mortality & - & 446 (89.62)  \\
\bottomrule
\end{tabular}
}
\end{table}
%\vspace{0.1cm}
%\footnotesize{Income proportion: Annual Household income level proportion; Manufacturing GDP: State GDP contributed by manufacturing; Obesity: Obesity among adults; Arthritis: Arthritis among adults; Current Smoking: Current cigarette smoking among adults;Premature death: years of potential life lost before age 75; Premature age-adjusted mortality: Number of deaths among residents under age 75 per 100,000 population (age-adjusted); Air pollution: Average daily density of fine particulate matter in micrograms per cubic meter (PM2.5); Poor physical health days: Average number of physically unhealthy days reported in past 30 days (age-adjusted)}

{These data provide a suitable setting for evaluating spatial heterogeneity in compositional socioeconomic factors, particularly in identifying regions with similar underlying structures that may not be geographically contiguous.}

\section{Methodology}\label{sec:method}

In this section, we introduce a Geographically Weighted Penalized Regression model designed specifically to account for spatial variations in regression relationships. Additionally, we address the challenges posed by compositional data, which include components such as proportions, percentages, or relative abundances that sum to a constant, typically 100\% or one. This characteristic complicates the application of traditional statistical models that typically assume

To effectively handle compositional data, we utilize a log-contrast regression model. This approach transforms compositional predictors into an unconstrained Euclidean space, preserving the meaningful relationships among the components. Consider a dataset with  $n$ independent observations, structured as  $(\mathbf{Y}, \tilde{\mathbf{X}}_{1}, \mathbf{X}_{2}) $, where $\mathbf{Y}=(y_1,\ldots,y_n)^\top$ represents the outcome vector,  $\tilde{\mathbf{X}}_{1}=(\tilde{\bm{x}}^\top_{11},\ldots, \tilde{\bm{x}}^\top_{1n})^\top$ is $n\times p$  compositional predictors matrix, and $\mathbf{X}_{2}=({\bm{x}}^\top_{21},\ldots, {\bm{x}}^\top_{2n})^\top$ is a  $n\times q$ non-compositional predictors matrix.

The model is specified as follows:
$$
\mathbf{Y} = \tilde{\mathbf{X}}_{1} \bar{\boldsymbol{\beta}} + \mathbf{X}_{2} \boldsymbol{\eta} + \boldsymbol{\epsilon},
$$
where \( \bar{\boldsymbol{\beta}} \) denotes the regression coefficients associated with the compositional predictors, \( \boldsymbol{\eta} \) represents the coefficients for the non-compositional predictors, and \( \boldsymbol{\epsilon} \) is the error term. This formulation allows us to analyze the effects of both compositional and non-compositional variables on the outcome.

Direct regression on \(\tilde{\mathbf{X}}_{1}\) is not feasible due to its compositional nature, and a common but naive solution is to exclude one component of the compositional vector from the regression analysis. However, this method introduces a dependency on the excluded component, resulting in inconsistencies in prediction and variable selection, thereby complicating model interpretation and inference. Following \citet{aitchison_statistical_1982,lin2014variable}, we address this issue by applying a log transformation to the compositional data, thereby converting it into a space that adheres to the familiar Euclidean geometry in \(\mathcal{R}^{p}\). Specifically, for each \(i = 1,\ldots,n\), we define \(z_{ij} = \log \tilde{x}_{1ij} \) for $j=1,\ldots p$, and represent \(\mathbf{Z}=[z_{ij}]\) as a matrix in \(\mathbb{R}^{n \times p}\). A small positive constant will be added only when $\tilde{x}_{1ij} = 0$ to avoid undefined values in the log transformation. This minimizes distortion in the overall composition compared to adding small positive constant to all values. This transformation allows us to reformulate the regression model as:
$$
\mathbf{Y}=\mathbf{Z}\tilde{\bm{\beta}}+\mathbf{X}_{2}\bm{\eta}+\bm{\epsilon}, \qquad  \text{s.t.}\quad \sum_{j=1}^p\tilde{\beta}_{j}=0,
$$
where \(\tilde{\bm{\beta}}=(\tilde{\beta}_{1},...,\tilde{\beta}_{p})^\top\) is the newly introduced coefficient vector for the transformed design matrix \(\mathbf{Z}\). We do not include an intercept in the model, as it can be eliminated by centering both the response $\mathbf{Y}$ and transformed design matrix $\mathbf{Z}$. Compared to the log-ratio transformation, this formulation is more convenient as it retains all compositional components in the regression model without requiring a reference component, thus preserving the full compositional structure while ensuring identifiability. {Unlike standard regression with categorical covariates that rely on a reference group, the compositional framework models relative changes among components under the sum-to-one constraint, thereby avoiding arbitrary reference selection.}

% how are 0 values handled in the log transformation?
% what is the new constraint that is imposed by the log transformation?
% doesnt make the last column of Z a constant?
% shouldn't it be:
% z_{i,j} = log(x_{i,j}/_x_{i,p}
% we use log transformation as lin 2014 and will add some small terms for zeros.
While the log transformation resolves the collinearity inherent in compositional data, it leads to regression coefficients that are not freely estimable due to the sum-to-zero constraint, which requires further transformations for proper analysis. To address this challenge, we implement a Helmert transformation, which projects the constrained coefficients onto an unconstrained space:
$$
\bm{\beta}=\mathbf{H}\tilde{\bm{\beta}},
$$
where \( \mathbf{H} \in \mathbb{R}^{(p-1) \times p} \) is the Helmert matrix with its first row omitted.

This transformation, while resolving the constraints of \(\tilde{\bm{\beta}}\), ensures that the regression coefficients remain interpretable and maintain the compositional data's geometric properties. We also utilize an orthogonal projection matrix, derived from a decomposition theorem, to ensure that the transformed coefficients \(\bm{\beta}\) can be accurately projected back to \(\tilde{\bm{\beta}}\), maintaining consistency with the inherent properties of compositional data.

\begin{theorem}
\label{the:orth_prior}
Let $\mathbf{H} = \mathbf{F}\mathbf{Q}$ be a full rank decomposition
of $\mathbf{H}$. Then $\mathbf{F}\in \mR^{(p-1)\times r}$ is a full column rank matrix and $\mathbf{Q}\in \mR^{r\times p}$ is a full row rank matrix. $\tilde{\mathbf{M}}=(\tilde{\mathbf{M}}_1,\tilde{\mathbf{M}}_2)=\mathbf{M}^{-1}$ with $\tilde\M_1\in \mR^{p\times (p-1)}$, where $\M = (\Q^\top,(\Q^\bot)^\top)^\top\in\mR^{p\times p}$ is the orthogonal compliment of $\Q$ satisfying $\Q(\Q^{\bot})^\top = \zero$, $\mathbf{Q}^{\bot}\in\mR^{(p-r)\times p}$ denotes the orthogonal compliment of $\mathbf{Q}$. Then the inverse transformation of $\mathbf{\tilde{\beta}}$ is 
\begin{equation*}
	\tilde{\bm{\beta}}=\tilde{\mathbf{M}}_1\bm{\beta}.
%\label{eq:fuse_prior}
\end{equation*}
%\label{the:inversetransform}
\end{theorem} 

This theorem enables the elimination of redundancy in the regression coefficients while preserving their geometric properties, and aligns with the methodology of \citet{egozcue_isometric_2003}, which established a robust framework for transforming constrained data while preserving geometric properties.

Applying Theorem~\ref{the:orth_prior}, we obtain the following unconstrained linear regression model:
\begin{equation}
\mathbf{Y}=\mathbf{X}_{1}\bm{\beta}+\mathbf{X}_{2}\bm{\eta}+\bm{\epsilon},
\label{eq:new regression}
\end{equation}
where $\mathbf{X}_1=\mathbf{Z}\tilde{\mathbf{M}}_1=(\bm{x}^\top_{11},\ldots,\bm{x}^\top_{1n})^\top$. Thus, through the log transformation and the Helmert matrix, we effectively remove the compositional constraint, making the regression model fully identifiable and interpretable in an unconstrained Euclidean space.

{For clarity, key notation used throughout the methodology is summarized in Table~\ref{tab:notation}.}

\begin{table}[tbp]
\centering
\renewcommand{\arraystretch}{1}
\setlength{\tabcolsep}{5pt}
\caption{{Summary of Notation}}
\label{tab:notation}
\begin{tabular}{ll}
\hline
Symbol & Description \\
\hline
$\mathbf{Y}$ & Response vector \\
$\mathbf{X}_1$ & Transformed compositional covariate matrix\\
$\mathbf{X}_2$ & Non-compositional covariate matrix\\
$\boldsymbol{\beta}$ & Coefficients for transformed compositional covariates \\
$\bm{\eta}$ & Coefficients for non-compositional covariates \\
$n$ & Number of spatial units \\
$p$ & Number of compositional components \\
%$\omega_{ij}$ & Spatial weight between locations $i$ and $j$ \\
%$d_{ij}$ & Distance between locations $i$ and $j$ \\
%$r$ & Spatial decay parameter controlling interaction range \\
%$\lambda$ & Regularization parameter for fusion penalty \\
%$p_\gamma(\cdot)$ & Non-convex penalty function (e.g., MCP) \\
\hline
\end{tabular}
\end{table}

Building on this foundation of transforming compositional data to enhance model interpretability, we can extend the analytical framework to incorporate spatial heterogeneity in the estimation process. Traditional regression approaches often ignore the spatial arrangement of data points, which can lead to unstable or inefficient coefficient estimates, especially in the presence of spatially structured data. It is essential to consider spatially varying compositional effects. These considerations are crucial in ensuring that the spatial characteristics of the data, which may influence the composition and its interaction with other variables, are adequately accounted for in the analysis. In the context of spatially varying compositional effects, we start with an understanding of how regression coefficients change with location, providing a granular view of local influences on the model's variables. However, while spatially varying compositional effects model is adept at highlighting local variability, it may overlook the broader spatial patterns that could inform more cohesive regional strategies or interventions. To address this, introducing a spatially aware penalization to encourage spatial clusters can significantly improve the model's performance. This method enforces similarity among regression coefficients of spatial units but traditionally applies a global penalty that does not differentiate based on geographic distance. Adapting a global pairwise penalty \citep{ma_exploration_2020} with geographical weights and model in \eqref{eq:new regression}, we propose following objective function for new penalized regression
\begin{equation}
    Q_n(\bm{\beta}, \bm{\eta}) = \frac{1}{2} \sum_{i=1}^n \left( y_i - \bm{x}_{1i} \bm{\beta}_i - \bm{x}_{2i} \bm{\eta} \right)^2 + \sum_{1 \leq i < j \leq n} p\left(\|\bm{\beta}_i - \bm{\beta}_j\|, \lambda, \omega_{ij} \right),
\label{eq:penalized_regression}
\end{equation}
where $\bm{\beta} = (\bm{\beta}_1^\top, \ldots, \bm{\beta}_n^\top)^\top$, and $p(\cdot, \lambda, \omega_{ij})$ is a geographically weighted concave penalty function redefined by the {MCP} or smoothly clipped absolute deviation (SCAD) \citep{fan_variable_2001} {. Here,} $\lambda \geq 0$ controls the strength of the penalty and $\omega$ represents geographical weights. %This method effectively clusters regions based on coefficient similarity and spatial proximity.This adjustment incorporates spatial priority into the regularization process, enhancing the model’s ability to reflect more realistic spatial relationships. 
{This formulation clusters regions based on coefficient similarity while incorporating spatial proximity.}The geographically weighted MCP can be expressed as:
\begin{equation*}
p_\gamma(t, \lambda,\omega_{ij}) = \omega_{ij} \lambda \int_0^t \left( 1 - \frac{x}{\gamma \lambda} \right)_+ dx, \quad \gamma > 1.
\label{eq: MCP}
\end{equation*}
{To further clarify the distinction between the proposed framework and existing approaches, we note that, unlike traditional SVCMs, which assume smooth spatial variation, or SCCMs, which rely on predefined partitions, the spatial weights {$\omega_{ij}$} govern the degree of similarity between locations: larger values encourage similar coefficients, whereas smaller values permit differences, thereby enabling adaptive clustering of spatial units.} %{\sout{We modified traditional penalty function to include a spatial weight term $\omega_{ij}$, which adjusts the strength of penalization based on geographical distance. The inclusion of geographical weights in the pairwise fusion penalty significantly enhances the model's ability to reflect the true spatial structure in the data.}}
By modulating the strength of the penalization based on geographic distances, { $\omega_{ij}$ encourages stronger similarity among nearby regions while allowing greater flexibility for distant ones. This mechanism helps balance over-smoothing, where true regional differences may be obscured, and under-smoothing, where important spatial dependencies may be ignored.} %\sout{the weights ensure that regression coefficients for closely located spatial units are more strongly encouraged to be similar compared to those that are further apart. This spatially sensitive approach helps to maintain a balance between over-smoothing, where too much similarity could mask true regional differences, and under-smoothing, where too little could ignore important spatial dependencies.}
%Furthermore, these weights effectively create a flexible framework that adapts the penalty not only based on the presence of coefficients but also based on their spatial context. This adaptation allows for a more detailed and nuanced understanding of spatial patterns, as it can dynamically adjust to varying degrees of spatial autocorrelation exhibited across different regions. As a result, the model can capture both global trends and local idiosyncrasies, providing a comprehensive analysis that is robust against spatial anomalies and discontinuities.

Comparing the existing geographically weighted regression
literature, our weights are obtained by graph distance between different areas. Following \citet{xue2020geographically,geng2022bayesian},
a graph $ G = (V, E) $ is defined, where $ V(G) = \{v_1, \dots, v_N\} $ represents the set of $ N $ vertices (e.g., spatial units such as states or counties), and $ E(G) = \{e_1, \dots, e_M\} $ denotes the set of $ M $ edges that represent connections between these units. The geographic distance between two vertices $v_i$ and $v_j$ is defined as:
\begin{equation*}
d_{v_i, v_j} =\min(|V(e)|),
\end{equation*}
where $|V(e)|$ denotes the number of target-level spatial unit along the shortest path $e$. {Each spatial unit is represented as a single vertex in the graph, regardless of its geographic shape or internal structure. It is important to distinguish this graph-based representation from geometric representations of spatial units, where a region may consist of multiple coordinates or polygon vertices. In this work, the graph is constructed at the level of areal units, with edges defined based on adjacency or distance between these units.} This geographic distance matrix is a matrix $n \times n$, with every entry representing pairwise geographic distances, with diagonal elements zero. The practical calculation of $d_{v_i, v_j}$ depends on the boundary-based adjacency: Two regions are adjacent if they share a common boundary, forming a direct connection in the graph.

For higher-level spatial units (e.g., states), boundaries can be defined based on actual geographic borders. However, for smaller units (e.g., counties), boundary definitions may be less practical. In such cases, adjacency is determined based on proximity thresholds of centroid distances, ensuring meaningful spatial relationships. Unlike Euclidean~\citep{duda2001pattern} or great-circle distances~\citep{snyder1987map}, which measure only physical proximity, the graph-based approach captures connectivity in a spatial network. This is particularly useful when spatial relationships reflect socioeconomic or transportation links rather than purely geographic distances. Especially in this study, Hawaii and Alaska are geographically isolated, making direct distance metrics ineffective.
Instead, we define their connections explicitly (e.g., Hawaii connected to California, Alaska connected to Washington), ensuring meaningful spatial interactions. 

{An exponential decay function is used for spatial weights to mitigate the impact of large jumps in spatial distances and ensure smooth transitions across locations.} Specifically, the spatial weight is calculated as:
\begin{equation}
\omega_{ij} = \exp(-d_{v_i, v_j}/r),
\label{eq:normal_weight}
\end{equation}
{where $r$ is a decay parameter that controls the rate at which spatial influence decays with distance, thereby determining the spatial scale of interaction. As $r$ decreases, the weights decay more rapidly, emphasizing local interactions and reducing the influence of distant regions. In contrast, larger values of $r$ lead to slower decay, allowing broader spatial interactions and incorporating more global information. In the limiting cases, as $r \to 0$, the model effectively captures only highly localized relationships, whereas as $r \to \infty$, the spatial weights approach uniformity, in which case the model reduces to pairwise fusion model of \citet{ma_exploration_2020}.}

{The spatial weights $\omega_{ij}$ govern the degree of similarity imposed between locations. Larger weights encourage greater similarity among coefficients, whereas smaller weights allow more pronounced differences. This mechanism provides a balance between over-smoothing, which may impose excessive similarity across regions, and under-smoothing, which may fail to capture meaningful spatial dependence.}

{While the exponential decay function in Equation~\eqref{eq:normal_weight} captures distance-based interactions, it may underweight directly adjacent locations, weakening local spatial dependence. To improve the balance between local and global spatial structure, we introduce an adjusted weighting scheme that strengthens local connections while preserving long-range interactions as: }
\begin{equation}
\omega_{ij} =
\begin{cases} 
    1, & \text{if } d_{v_i, v_j} = 1, \\
    \exp(-d_{v_i, v_j}/r), & \text{if } d_{v_i, v_j} > 1.
\end{cases}
\\[1em]
\label{eq:target_weight}
\end{equation}

\begin{figure}[tbp]
    \centering
    \includegraphics[width=6in]{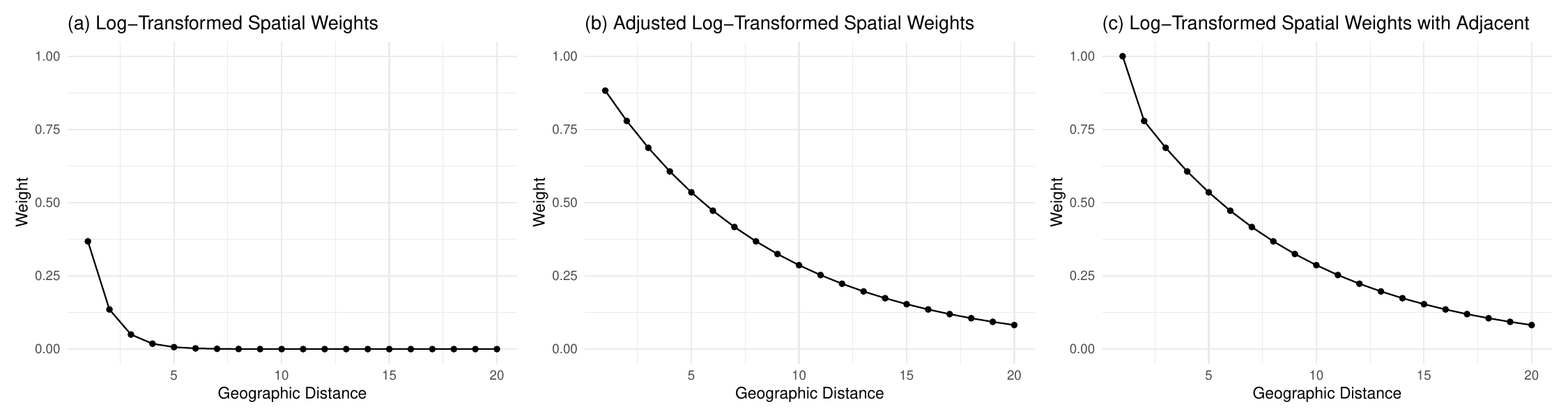}
    \caption{Spatial weight patterns under different method, with r=8 for (b) and (c)}
    \label{fig: Weighting comparison}
\end{figure}

Figure~\ref{fig: Weighting comparison} illustrates the differences between the original exponential weights, the adjusted weighting scheme, and adjacency-based approaches. {The comparison highlights how the proposed adjustment increases the influence of nearby regions while maintaining a smooth decay over longer distances, resulting in a more balanced representation of spatial dependence across the spatial effect range.}

{The choice of $r$ mainly affects the granularity of the detected spatial clusters rather than the underlying structure. Smaller values of $r$ emphasize local differences and therefore tend to produce finer clusters, whereas larger values lead to more aggregated spatial regimes. These differences are best understood as changes in resolution rather than as evidence of fundamentally different clustering patterns. In practice, clusters obtained with smaller values of $r$ are often interpretable as subdivisions of those identified with larger values, indicating a hierarchical structure rather than instability.The impact of the decay parameter $r$ on the resulting spatial clustering is further examined through a sensitivity analysis, presented in Section~\ref{sec:sensitivity}, to assess the robustness of the identified spatial structures.}

To explore the impact of different levels of penalization, we estimate $\hat{\bm{\eta}}(\lambda)$ and $\hat{\bm{\beta}}(\lambda)$ over a range of values, defined as $[\lambda_{\text{min}}, \lambda_{\text{max}}]$. The maximum value, $\lambda_{\text{max}}$, forces all coefficients to become identical, effectively eliminating heterogeneity. The minimum value, $\lambda_{\text{min}}$, is a small positive number that allows flexibility in clustering solutions. The estimated solution path $\hat{\bm{\beta}}(\lambda)$ is examined in this range to identify stable clustering patterns. Estimation is performed using the ADMM algorithm, a powerful optimization technique well suited to solving large-scale penalized regression problems. ADMM decomposes the global optimization problem into manageable subproblems, ensuring theoretical convergence under concave penalties. This approach enables efficient estimation of parameters while maintaining computational feasibility for large datasets.

\begin{algorithm}[tbp]
\caption{Computational Algorithm using ADMM}
\label{alg:admm}
\resizebox{7in}{!}{%
\begin{minipage}{7.7in}
\begin{algorithmic}[1]
\REQUIRE Initialize parameters: $\bm{\beta}$, $\bm{\eta}$, $\bm{\delta}$, $\lambda$, and $\bm{\alpha}$ using Appendix Section B.
\WHILE{convergence criterion is not met}
    \STATE Update $\bm{\beta}^{m+1}$ to minimize the regression error.
    \STATE Update $\bm{\eta}^{m+1}$ to refine the effect of non-compositional covariates.
    \STATE Apply spatial penalty and adjust $\bm{\delta}^{m+1}$ accordingly.
    \STATE Update $\bm{\alpha}^{m+1}$ for auxiliary constraints.
    \STATE Increment iteration: $m = m + 1$.
\ENDWHILE 
\STATE Denote the final solution as $(\hat{\bm\eta}(\lambda,\omega), \hat{\bm\beta}(\lambda,\omega))$.
\ENSURE Final output: $(\hat{\bm\eta}(\lambda,\omega), \hat{\bm\beta}(\lambda,\omega))$.
\end{algorithmic}
\end{minipage}
}
\end{algorithm}

The algorithm~\ref{alg:admm} outlines the steps of the ADMM procedure, including initialization, iterative updates, and convergence criteria. For detailed implementation, including initialization, update rules, and stopping criteria, refer to Appendix Section A, where the complete algorithm is described.

In addition to spatial weight, the parameter $\lambda$ plays a critical role in balancing the sparsity and clustering resolution. A data-driven selection procedure based on a modified Bayesian Information Criterion \citep[Modified BIC]{ma_exploration_2020} is applied to determine the optimal value of $\lambda$. 
The modified BIC is defined as:
\begin{align}
\text{BIC}_\text{modified}(\lambda) = \log \left( \frac{1}{n} \sum_{i=1}^n \left( y_i - \bm{x}_{1i} \hat{\bm{\beta}_i}(\lambda) - \bm{x}_{2i}\hat{\bm{\eta}}(\lambda) \right)^2 \right) 
+  C_n\frac{\log n}{n} \left( \hat{K}(\lambda)p + q \right), \label{eq:bic}
\end{align}
where $C_n$ is a positive number and we define $C_n = \log(np + q)$, this modified BIC would reduces to traditional Bayesian Information Criterion  \citep[BIC]{wang_tuning_2007} when $C_n = 1$. $\hat{K}(\lambda)$ is the number of estimated clusters corresponding to unique values of $\hat{\beta}$.
The optimal $\lambda$ is selected by minimizing the modified $\text{BIC}(\lambda)$:
\begin{equation*} 
\hat{\lambda} = \arg \min_{\lambda} \text{BIC}(\lambda). 
\end{equation*}
% For simplicity, we define: 
% \begin{equation*} 
% (\hat{\bm\beta}, \hat{\bm\eta}) \equiv (\hat{\bm\beta}(\hat{\lambda}), \hat{\bm\eta}(\hat{\lambda})).
% \end{equation*} 

\section{Simulation studies}\label{sec:simulation}

\subsection{\textbf{\textit{Simulation settings}}}\label{sec:simSettings}

We implement a simulation framework in which the tuning parameter is selected using the {Modified BIC} defined in Equation~\eqref{eq:bic}. The simulation employs {minimax concave penalty (MCP)} to promote the grouping of spatial units with similar coefficients while mitigating the risk of over-shrinkage. The experimental design incorporates realistic spatial heterogeneity and compositional dependencies, along with controlled lattice-based configurations for benchmarking. For each setting, we {assess the ability of the method} to recover the true spatial cluster structure {and estimate the associated coefficients accurately}.

{We compare the proposed method with several benchmark approaches, including a constant model without spatial structure, an adjacency-based clustering method that enforces strict spatial contiguity, and a spatial pairwise fusion model without geographically weighted adjustments. Detailed formulations of these methods are provided in Section~\ref{sec:method}.}

{To evaluate the proposed framework in spatial settings relevant to geographic applications, we construct synthetic datasets that capture both spatial heterogeneity and compositional dependence.} The simulations encompass configurations with 51 spatial units (U.S. states) and 254 spatial units (Texas counties), representing settings with varying degrees of spatial granularity. These units are partitioned into distinct clusters according to predefined spatial structures, thereby preserving realistic spatial dependence patterns. 

\begin{figure}[htbp]
    \centering
    \begin{subfigure}
        \centering
        \includegraphics[width=5.5in]{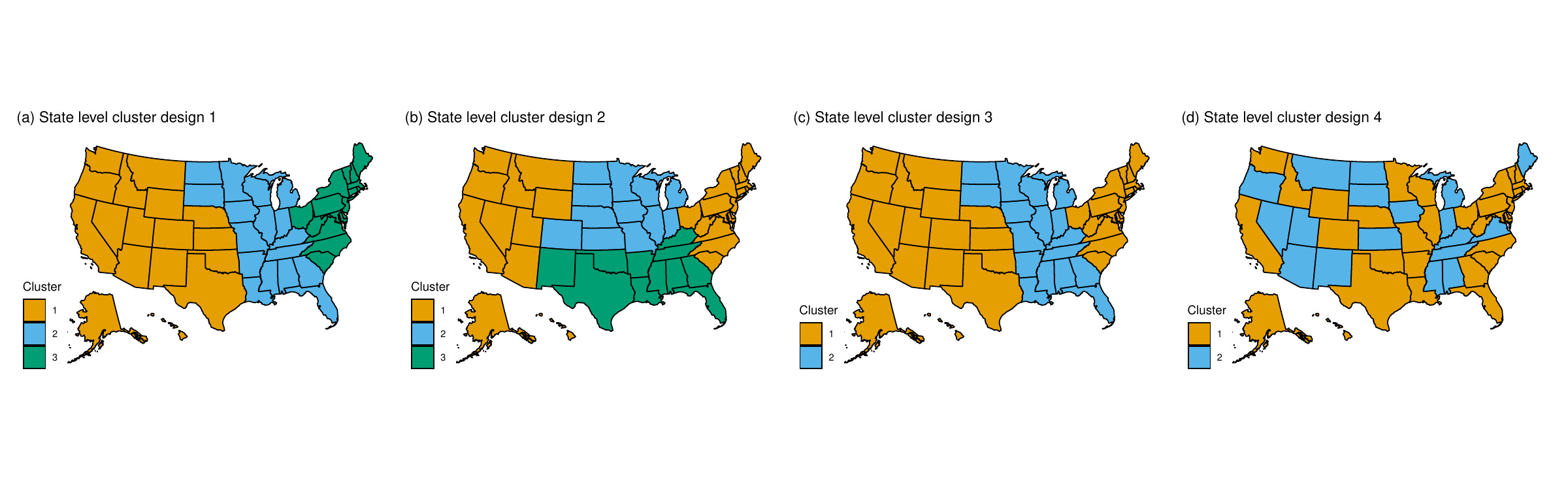}
    \end{subfigure}
    \hfill
    \begin{subfigure}
        \centering
        \includegraphics[width=5.5in]{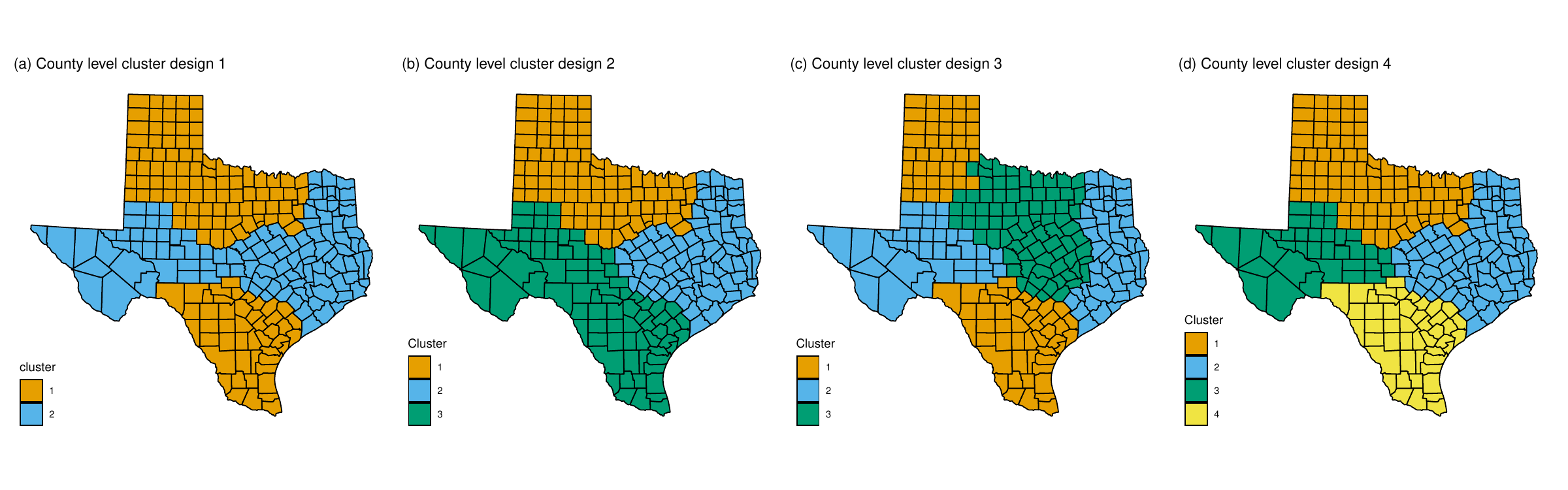}
    \end{subfigure}
    \caption{Spatial partitions at different geographic resolutions. The upper figure shows the state-level design and the middle figure shows the Texas county-level design}
    \label{fig:combined_simmaps}
\end{figure}

{The spatial partition designs used in the simulations, illustrated in Figures~\ref{fig:combined_simmaps}, are inspired by commonly observed geographic patterns and represent different clustering structures. Specifically, at the state level, Design 1 reflects broad regional divisions such as the Western, Central, and Eastern United States, capturing large-scale spatial heterogeneity; Design 2 design represents variation along socioeconomic or urban–rural gradients, where regions with similar levels of development or population density are grouped together regardless of strict geographic proximity; Design 3 groups coastal regions into the same cluster to reflect similarities in economic structure, urbanization, and health-related characteristics despite geographic separation; and Design 4 design imposes weaker spatial constraints, allowing for more flexible grouping structures that deviate from strict geographic contiguity. A similar design principle is adopted at the county level, with the goal of capturing realistic geographic clustering patterns beyond simple spatial proximity. At the county level, spatial partitions are constructed based on the Public Health Regions defined by the Texas Department of State Health Services (DSHS) \citep{dshs_regions}. These regions are widely used in public health practice to capture similarities in healthcare infrastructure, demographic composition, and disease patterns. As such, they provide a meaningful geographic basis for grouping counties into clusters that reflect real-world epidemiological and administrative structures.}

{In addition, a lattice-based spatial structure is included in the Appendix to provide a controlled experimental setting in which spatial relationships are explicitly defined through regular adjacency. The lattice design offers a fundamental benchmark for validating, testing, and comparing model performance under systematically structured spatial dependence. A lattice simulation is needed because it provides a transparent, standardized benchmark with known spatial structure, making it easier to isolate and evaluate the method’s ability to recover spatial dependence and clustering patterns.}

The compositional predictors $\bm{x}_{0i}$ are generated as three-dimensional vectors ($p = 3$) constrained to sum to one. Each element of $\bm{x}_{0i}$ is initially drawn from a uniform distribution and subsequently normalized by its row sum. A logarithmic transformation followed by a Helmert transformation is applied to obtain an orthonormal representation of the compositional components suitable for regression analysis. In addition, non-compositional predictors $\bm{x}_{2i}$ are generated independently as two-dimensional vectors ($q = 2$), with each element sampled from a uniform distribution.

The outcome variable $y$ is generated as
\begin{align*}
\bm{Y} = \bm{X}_1 \bm{\beta} + \bm{X}_2 \bm{\eta} + \bm{\epsilon},
\end{align*}
where $\bm{X}_1$ denotes the transformed compositional matrix derived from $\bm{X}_0$, $\bm{\beta}$ represents the true coefficients assigned according to spatial cluster membership, and $\bm{\eta} = (1, 1)$ captures non-spatial effects. Random noise $\bm{\epsilon} \sim \mathcal{N}(0, \sigma^2)$ with $\sigma^2 = 0.1$ is added to account for observational variability.

Spatial dependence is explicitly incorporated into the data-generating process such that spatial units within the same cluster share similar coefficients $\bm{\beta}$. These latent clusters form the ground truth for evaluating the proposed model’s ability to recover both spatial structure and coefficient patterns accurately.

To comprehensively evaluate model performance, we assess both \textit{clustering accuracy} and \textit{estimation accuracy} metrics.

{For clustering performance, we consider three commonly used metrics: Rand Index (RI) \citep{rand_objective_1971}, Clustering Accuracy (CA), and Relative Cluster Count (RCC). RI evaluates pairwise clustering agreement, with values closer to 1 indicating better recovery of the true clustering structure. CA measures the proportion of correctly assigned spatial units after aligning estimated labels with true labels, where higher values indicate better performance. RCC assesses how well the estimated number of clusters matches the true number. An RCC value close to 1 indicates accurate recovery, whereas values greater than 1 suggest over-clustering and values less than 1 indicate under-clustering.}

To evaluate estimation accuracy, we report bias and mean squared error (MSE) of the regression coefficients. Bias reflects the average deviation from the true coefficients, while MSE captures both variance and squared bias. Lower values of both metrics indicate better estimation performance. 

Detailed definitions and formulas for evaluation metrics are provided in the Appendix.

\subsection{\textbf{\textit{US state level simulation}}}\label{sec:ussimulation}

We designed four cluster partitions for the state-level simulation, each reflecting different spatial structures that could arise in various research contexts. These designs are illustrated in Figure~\ref{fig:combined_simmaps} upper figure. These designs vary in contiguity, number of clusters, and separation strength, ranging from conventional regional division to noncontiguous groupings and simplified or highly distinct structures.Together, they provide a comprehensive setting for simulation. 

{The results in Table~\ref{tab:state_summary} show that incorporating spatial information substantially improves both clustering and estimation accuracy. The proposed \textit{Spatial-adjacency} method consistently achieves the highest or near-highest RI and CA values across all designs, while maintaining RCC values close to 1, indicating accurate recovery of the true cluster structure without systematic over- or under-clustering. 
In contrast, the \textit{Pure adjacency} method tends to over-cluster, whereas the \textit{Constant} method often under-clusters. The \textit{Spatial-pairwise} method provides moderate improvement but remains less stable in balancing cluster granularity and consistency. Overall, the proposed method achieves a better balance between cluster recovery and model stability across different spatial configurations.}

\renewcommand{\thetable}{\arabic{table}}
\begin{table}[tbp]
\centering
\caption {\label{tab:state_summary} Summary of state-level simulation performance} 
\vspace{0.1cm}
\resizebox{5in}{!}{%
\begin{tabular}{lllrrrrrrr}
\toprule
 &  &  & \multicolumn{3}{l}{Performance} & \multicolumn{2}{l}{Estimation-$\beta$} & \multicolumn{2}{l}{Estimation-$\eta$} \\ \cmidrule(lr){4-10}
\multicolumn{1}{l}{Design} & Method       & r    & RI   & CA   & RCC   & Bias & MSE  & Bias & MSE  \\ 
\midrule
\multicolumn{1}{c}{1}   & Constant     & -    & 0.72 & 0.64 & 0.69  & 0.30 & 0.15 & 0.15 & 0.22 \\
\multicolumn{1}{l}{}          & Pure adjacency       & -    & 0.72 & 0.27 & 6.38  & 0.11 & 0.03 & 0.14 & 0.03 \\
\multicolumn{1}{l}{}          & Spatial-pairwise      & 0.8 & 0.73 & 0.64 & 1.07  & 0.26 & 0.32 & 0.24 & 0.13 \\
\multicolumn{1}{l}{}          & Spatial-adjacency & 0.8 & \textbf{0.75} & \textbf{0.67} & 1.18 & \textbf{0.22} & \textbf{0.22} & 0.19 & 0.05 \\
\midrule
\multicolumn{1}{c}{2}   & Constant     & -    & 0.69 & 0.71 & 0.71  & 0.33 & 0.22 & 0.40 & 0.23 \\
\multicolumn{1}{l}{}          & Pure adjacency       & -    & 0.65 & 0.25 & 6.67  & 0.04 & 0.00 & 0.11 & 0.01 \\
\multicolumn{1}{l}{}          & Spatial-pairwise      & 2.0    & 0.70 & 0.72 & 0.80  & \textbf{0.29} & \textbf{0.15} & 0.35 & 0.17 \\
\multicolumn{1}{l}{}          & Spatial-adjacency & 3.5  & \textbf{0.70} & \textbf{0.72} & 0.76  & 0.30 & 0.16 & 0.37 & 0.18 \\
\midrule
\multicolumn{1}{c}{3}   & Constant     & -    & 0.66 & 0.77 & 0.83  & 0.21 & 0.09 & 0.13 & 0.03 \\
\multicolumn{1}{l}{}          & Pure adjacency       & -    & 0.49 & 0.19 & 10.65 & 0.03 & 0.00 & 0.10 & 0.01 \\
\multicolumn{1}{l}{}          & Spatial-pairwise      & 2.5  & 0.75 & 0.83 & 1.05  & \textbf{0.10} & \textbf{0.02} & 0.11 & 0.02 \\
\multicolumn{1}{l}{}          & Spatial-adjacency & 3.5  & \textbf{0.75} & \textbf{0.85} & 1.03  & \textbf{0.10} & \textbf{0.02} & 0.10 & 0.02 \\
\midrule
\multicolumn{1}{c}{4}   & Constant     & -    & 0.65 & 0.76 & 0.93  & 0.18 & 0.08 & 0.13 & 0.03 \\
\multicolumn{1}{l}{}          & Pure adjacency       & -    & 0.48 & 0.14 & 12.3  & 0.02 & 0.00 & 0.13 & 0.02 \\
\multicolumn{1}{l}{}          & Spatial-pairwise      & 2.0    & 0.69 & \textbf{0.79 }& 1.00  & 0.12 & 0.03 & 0.12 & 0.02 \\
\multicolumn{1}{l}{}          & Spatial-adjacency & 2.5  & \textbf{0.69} & 0.75 & 1.20  & \textbf{0.09} & \textbf{0.01} & 0.17 & 0.04\\
\bottomrule
\end{tabular}
}
\end{table}

Table~\ref{tab:state_summary} also shows that the target method \textit{Spatial-adjacency} produces estimated values of compositional coefficient $\bm{\beta}$ closely approximate the true values, with consistently low bias and MSE across all designs. Although Design 2 exhibits slightly higher values, the estimate remains within an acceptable range. While in Design 1 and Design 2, the bias and MSE of non-compositional coefficient $\bm{\eta}$ are not the lowest among all methods, their estimate still low and indicating satisfactory performance.

\subsection{\textbf{\textit{Texas county level simulation}}}

{To further evaluate the proposed method at a finer spatial resolution}, we conducted simulations at the Texas county level, {which involve a larger number of spatial units and more complex spatial configurations}. This setting allows us to assess the model’s scalability and effectiveness in handling a larger number of spatial units while maintaining the compositional structure of the predictors. The four clustering designs for the Texas county-level simulation are illustrated in Figure~\ref{fig:combined_simmaps}.{The simulation setup follows the same data-generating process as described in Subsection~\ref{sec:simSettings}, ensuring comparability with the state-level analysis.}

The results in Table~\ref{tab:county_summary} show that the proposed \textit{Spatial-adjacency} method consistently achieves the highest or near-highest RI and CA values across all designs, with RCC remaining close to 1. Although RI rounded to two decimal places, the actual values are slightly higher than those of the \textit{Spatial-pairwise} method, confirming a consistent improvement. These findings {are consistent with state-level simulation results} that target method is able to accurately cover both cluster membership and cluster number, even in higher-resolution settings. Notably, \textit{Pure adjacency} method failed to converge due to the excessive sparsity of the adjacency matrix. For such a large sample size, the method performs poorly or becomes infeasible to run.

This indicates accurate recovery of both cluster membership and cluster number, even in higher-resolution settings.

\renewcommand{\thetable}{\arabic{table}}
\begin{table}[tbp]
\centering
\caption {\label{tab:county_summary} Summary of county-level simulation performance} 
\vspace{0.1cm}
\resizebox{5in}{!}{%
\begin{tabular}{lllrrrrrrr}
\toprule
 &  &  & \multicolumn{3}{l}{Performance} & \multicolumn{2}{l}{Estimation-$\beta$} & \multicolumn{2}{l}{Estimation-$\eta$} \\ \cmidrule(lr){4-10}
\multicolumn{1}{l}{Design} & Method       & r    & RI   & CA   & RCC   & Bias & MSE  & Bias & MSE  \\ 
\midrule
\multicolumn{1}{c}{1}      & Constant  & -    & 0.91 & 0.96 & 1 & 0.17 &  0.05 & 0.06  &  0.00 \\
\multicolumn{1}{l}{}          &  Pure adjacency & -    & - & - & - & - &  - & -  &  - \\
\multicolumn{1}{l}{} & Spatial-pairwise      & 2.0  & 0.93  & 0.96  & 1.05  &  \textbf{0.09}   & \textbf{0.01}    & 0.06     & 0.00     \\
\multicolumn{1}{l}{} & Spatial-adjacency & 2.0 & \textbf{0.93} & \textbf{0.96} & 1.23 & 0.10 & 0.02 & 0.05 & 0.00   \\
\midrule
 \multicolumn{1}{c}{2}      & Constant     & -    & 0.64  & 0.66  & 1.04   & 0.13  & 0.06  & 0.40  & 0.21  \\
\multicolumn{1}{l}{}          &  Pure adjacency       & -    & -  & -  & -  & - & -  & -  & -  \\
\multicolumn{1}{l}{} & Spatial-pairwise      & 2.5 & 0.69 & \textbf{0.70} & 1.08 & 0.11 & \textbf{0.02}    & 0.30             & 0.12            \\
\multicolumn{1}{l}{} & Spatial-adjacency & 2.5 & \textbf{0.69} & 0.69  & 1.09 & \textbf{0.10}    & \textbf{0.02}    & 0.27             & 0.10            \\ 
\midrule
\multicolumn{1}{c}{3}      & Constant     & -    & 0.58 & 0.51 & 0.83  & 0.45 & 0.36 & 0.21 & 0.06 \\
\multicolumn{1}{l}{}       &  Pure adjacency       & -    & - & - & - & - & -    & -     & -     \\
\multicolumn{1}{l}{} & Spatial-pairwise  & 2.5  & 0.62 & \textbf{0.46} & 1.13 &  \textbf{0.22}  & 0.11 & 0.13  & 0.02 \\
\multicolumn{1}{l}{} & Spatial-adjacency & 3.0 & \textbf{0.63} & 0.45 & 1.11 & 0.23 & \textbf{0.10} & 0.13 & 0.02 \\
\midrule
\multicolumn{1}{c}{4}      & Constant     & -    & 0.69 & 0.59 & 0.72  & 0.91 & 1.56 & 0.40 & 0.21 \\
\multicolumn{1}{l}{}          & Pure adjacency       & -    & - & - & -  & - & - & - & - \\
\multicolumn{1}{l}{}          & Spatial-pairwise  &  2.0    &  0.71     &  0.62     &    0.85    &    \textbf{0.52}   &   \textbf{0.56}    &   0.30    &    0.11   \\
\multicolumn{1}{l}{}          & Spatial-adjacency & 1.0 & \textbf{0.71} & \textbf{0.63} & 1.00  & 0.58 & 0.86 & 0.34 & 0.14\\
\bottomrule
\end{tabular}
}
\end{table}

The coefficient estimation result in Table~\ref{tab:county_summary} highlights the superior estimation precision of the proposed \textit{spatial-adjacency} method, particularly for the compositional coefficient $\bm{\beta}$. Across designs, \textit{Spatial-adjacency} achieves the lowest bias and MSE in most case. For the non-compositional coefficient $\bm{\eta}$, our target method also demonstrates strong performance, achieving competitive low bias and MSE values. While the \textit{Spatial-pairwise} method occasionally attains slightly lower MSE for $\bm{\eta}$, its overall performance remains less stable. Taken together, these results underscore the effectiveness of the \textit{Spatial-adjacency} approach in delivering accurate and reliable parameter estimates, while simultaneously preserving spatial structure and clustering coherence in complex county-level settings.

{
Additional sensitivity checks indicate that the main clustering patterns remain consistent across a range of $r$ values in both simulation settings, with differences primarily reflecting changes in resolution rather than structural instability.
Overall, the simulation results demonstrate that the proposed Spatial-adjacency method consistently improves both clustering accuracy and estimation performance across different spatial configurations and resolutions. In particular, the method effectively recovers spatial clusters, including non-contiguous regions with similar underlying structures, while maintaining stable estimation of regression coefficients.
Compared to existing approaches, the proposed method achieves a better balance between cluster recovery and model stability, avoiding both over-clustering and under-clustering behaviors observed in alternative methods. These advantages are particularly evident in settings with complex spatial structures, where traditional approaches based on strict adjacency or smoothness assumptions tend to perform poorly.
These findings highlight the importance of allowing spatial clusters to be defined by structural similarity rather than strict geographic proximity, which is essential for capturing realistic spatial patterns in geographic and public health applications.
}

%{Overall, the simulation results consistently demonstrate that the proposed method effectively identifies spatially heterogeneous regimes while maintaining robustness across different settings, with improvements primarily reflected in clustering accuracy and interpretability.}

\section{Sensitivity Analysis}\label{sec:sensitivity}

{While the simulation results demonstrate the overall performance of the proposed method, it is also important to assess the sensitivity of the clustering outcomes to the choice of tuning parameters. We therefore examine the effect of the decay parameter $r$ by conducting a sensitivity analysis over a range of representative values using simulated data based on the state-level design 1 (Figure~\ref{fig:combined_simmaps}).}

\begin{figure}
    \centering
    \includegraphics[width=1\linewidth]{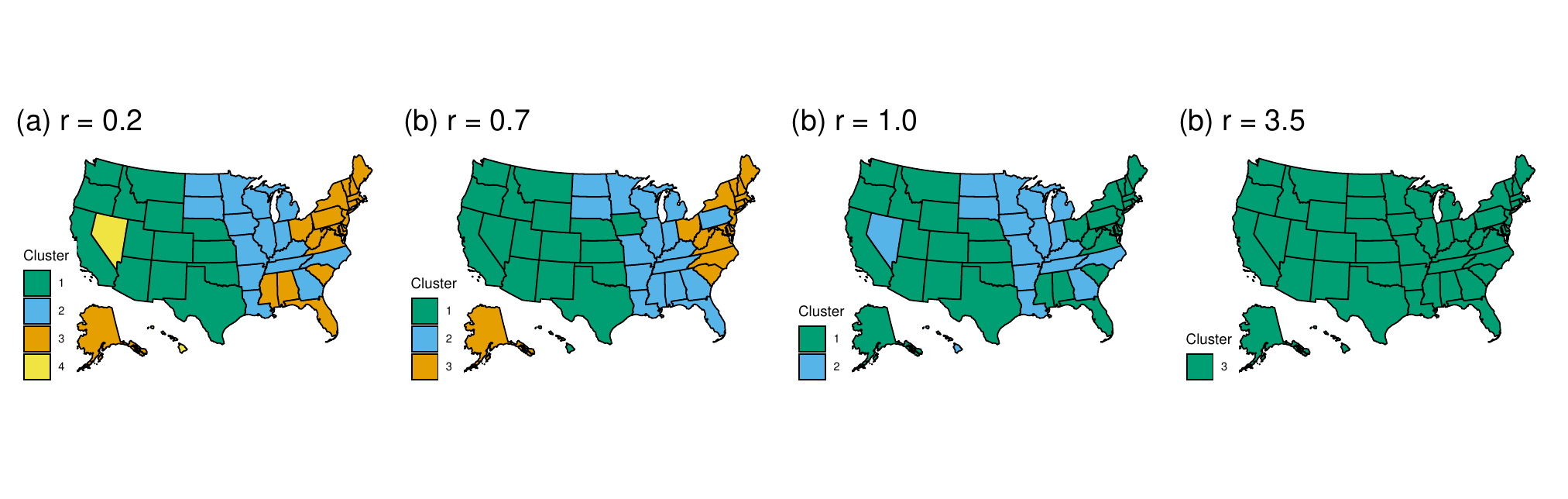}
    \caption{Spatial clustering results under different values of the decay parameter $r$.}
    \label{fig:sensitivity_r}
\end{figure}

{Figure~\ref{fig:sensitivity_r} presents the clustering results for different values of $r$. As $r$ increases, the number of detected clusters decreases and the spatial regimes become progressively more aggregated. For sufficiently large values of $r$ (e.g., $r=3.5$), the model yields a single dominant cluster.}

{Importantly, the overall spatial structure remains largely consistent across different values of $r$. The observed differences are primarily reflected in the level of granularity rather than in fundamentally different clustering patterns. In particular, clusters identified under smaller values of $r$ can often be interpreted as subdivisions of those obtained under larger values, suggesting an underlying hierarchical spatial organization rather than instability of the method.}

{These results suggest that the proposed method is robust to the choice of $r$, with the parameter primarily serving as a tuning mechanism to control the resolution of spatial clustering.}

\section{Real data analysis}\label{sec:real_data}

\subsection{\textbf{\textit{Income categorization and population adjustment}}}

Compositional variables of 2022 annual household income for both state-level and county-level originally categorized into 16 and 11 income brackets, were aggregated into four categories to ensure comparability and interpretability across regions. The categorization aligns with widely used definitions of income tiers based on the U.S. median income, where lower-income households earn less than two-thirds of the median and upper-income households earn more than double the median \citep{pew2024middleclass}. Specifically:
\begin{itemize}
    \item For U.S. states: low ($<$ \$30,000), low-medium ($<$ \$50,000), medium ($<$ \$100,000), and high income ($>=$ \$100,000).
    \item For Texas counties: low ($<$ \$25,000), low-medium ($<$ \$50,000), medium ($<$ \$100,000), and high income ($>=$ \$100,000).
\end{itemize}
The slightly modified categorization for Texas counties reflects differences in the available data granularity between data sources.

At county level, the model incorporated adjustments to account for population density disparities, ensuring that counties with small populations did not disproportionately influence clustering results. In particular, some Texas counties have smaller sample sizes, leading to higher variance in their estimated COPD prevalence. To address this issue, we applied a population adjustment to stabilize variance and enhance the reliability of the clustering process.

\subsection{\textbf{\textit{U.S. states level results}}}

The analysis was conducted to examine the relationship between COPD prevalence and income distribution across states or counties.

\begin{figure}[tbp]
    \centering
    \includegraphics[width=5.5in]{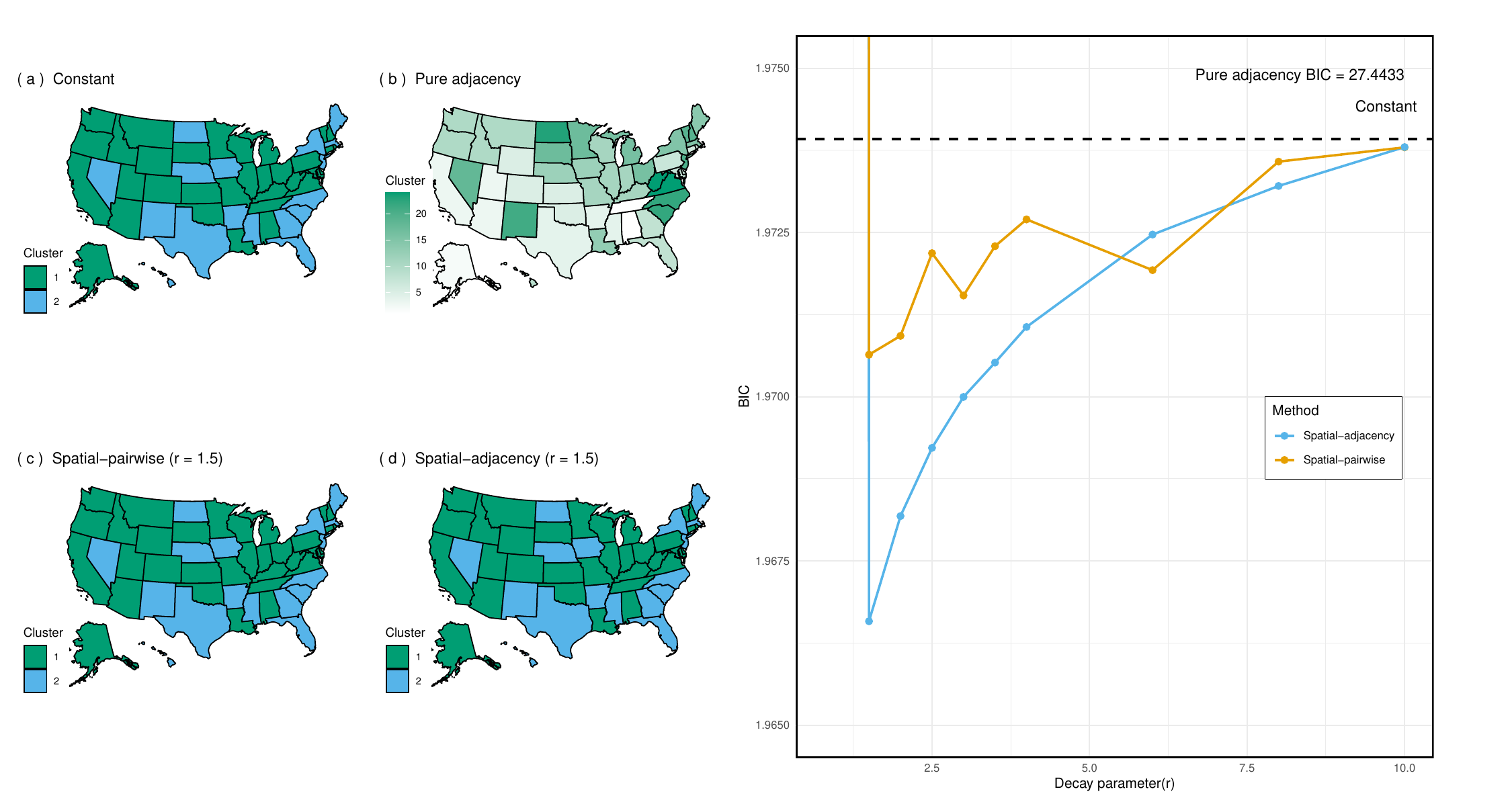}
    \caption{State-level clustering results (left) and BIC comparison (right) under different spatial weighting methods.
    }
    \label{fig:usrealresult}
\end{figure}

We evaluated the performance of four different weighting methods by both visualizing the clustering results on the US map and comparing the BIC values, as shown in Figure~\ref{fig:usrealresult}. {
The BIC comparison reveals clear differences in model fit across methods. In particular, the proposed spatial weighting approach achieves lowest modified BIC values, indicating improved model performance and better recovery of the underlying spatial structure. The optimal value of $r$ identified by BIC further supports the importance of appropriately balancing local and global spatial interactions. Although the clustering patterns may appear visually similar across methods, the resulting partition maps reveal meaningful regional structures. In the real data analysis, except for the pure adjacency model, which tends to over-cluster, the remaining three methods produce coherent and interpretable spatial regimes.
}

In the real data analysis, except pure adjacency model highly over clustered, rest three methods produced coherent and meaningful clusters. The resulting partition map revealed meaningful regional patterns that aligned with expected spatial dependencies. On the model selection side, the BIC comparison showed that spatial-adjacency achieved the lowest values across a range of decay parameters with a BIC = 1.967 at r = 1.5, indicating superior model fit relative to competing methods. The spatial-pairwise method exhibited greater variability in BIC and consistently underperformed, while the constant and pure adjacency models yielded a much higher BIC values with lowest value of 1.970 at r = 1.5. These results highlight the advantage of incorporating adjacency-based spatial penalization for capturing latent spatial heterogeneity in real-world settings in {U.S. }state level setting. 

\begin{table}[htbp]
\centering
\caption{Estimate results of regression coefficients in state-level}
\label{tab:us_beta}
\setlength{\tabcolsep}{3.5pt}
\noindent\makebox[\textwidth]{ % Centering the resized table
\begin{tabular}{p{3cm}rr}
\toprule
& \multicolumn{2}{c}{\textbf{Spatial-adjacency}} \\
& \multicolumn{2}{c}{($\lambda = 0.58$)} \\
& \textbf{Cluster 1} & \textbf{Cluster 2} \\
\midrule
$\bm\beta_{\text{low}}$ & $0.23$ & $0.03$\\
$\bm\beta_{\text{low-med}}$ & $4.15$ & $5.94$\\
$\bm\beta_{\text{medium}}$ & $-4.39$ & $-7.17$\\
$\bm\beta_{\text{high}}$ & $0.01$ & $1.20$\\
\midrule
$\bm\eta_{\text{Manufacturing}}$ & \multicolumn{2}{c}{$-0.17$}\\
$\bm\eta_{\text{Adult obese}}$ & \multicolumn{2}{c}{$-0.11$}\\
$\bm\eta_{\text{Arthritis}}$ & \multicolumn{2}{c}{$-0.11$}\\
$\bm\eta_{\text{Smoking}}$ & \multicolumn{2}{c}{$-0.04$}\\
\bottomrule
\end{tabular}
}
\end{table}

{
To further interpret the estimated spatial relationships, Table~\ref{tab:us_beta} presents the regression coefficients from the state-level analysis under the spatial-adjacency method.
The compositional coefficients represent relative effects among income groups. 
In both clusters, the low- and lower-middle-income components exhibit consistent positive associations with COPD prevalence. This finding corroborates the well-documented geographic concentration of respiratory health disparities in socioeconomically disadvantaged areas \citep{Pleasants2016, grigsby2016socioeconomic}.
Notably, the medium-income component shows a strong negative association across both clusters, suggesting that a higher share of middle-income households is consistently associated with lower COPD prevalence. This pattern highlights a potential protective effect of middle-income composition in mitigating adverse health outcomes.
Interestingly, the high-income component ($\bm\beta_{\text{high}}$) exhibits heterogeneity: while its effect is negligible in Cluster 1, it is notably positive in Cluster 2. This pattern may be interpreted as a "wealth–prevalence paradox" attributed to diagnostic bias, as regions with higher high-income shares often possess superior medical infrastructure and higher screening rates, leading to increased reporting of chronic conditions \citep{Pleasants2016}. The estimated coefficients for non-compositional covariates are generally consistent across clusters. In particular, variables such as smoking and arthritis show negative associations with COPD prevalence in this analysis, suggesting that these factors may reflect underlying regional differences in health conditions or data reporting patterns rather than direct causal effects.
%The estimated coefficients for non-compositional covariates are relatively stable across clusters. While several variables exhibit negative associations with COPD prevalence in this analysis, these effects should be interpreted cautiously, as they may reflect underlying regional patterns, confounding factors, or differences in data aggregation rather than direct causal relationships.
Overall, these results demonstrate that income composition, rather than aggregate income levels alone, plays an important role in shaping spatial disparities in COPD prevalence. The variation in coefficient patterns across clusters further indicates that these relationships differ across regions, suggesting that these associations may reflect underlying regional patterns or differences in data aggregation, rather than direct causal effects.
}

{The detected clusters in Figure~\ref{fig:usrealresult} exhibit meaningful alignment with broad U.S. regional patterns, providing qualitative support for the identified spatial regimes. Cluster 1 largely includes states that share relatively stable socioeconomic structures and historical industrial bases, while Cluster 2 captures regions with more heterogeneous and rapidly evolving socioeconomic conditions. Some non-contiguous regions are grouped together due to similarities in underlying socioeconomic and health-related characteristics, rather than geographic proximity alone. This highlights the ability of the proposed method to identify spatial regimes driven by structural similarity rather than predefined administrative boundaries. The observed differences in $\bm\beta_{\text{high}}$ across clusters further suggest that the relationship between income composition and COPD prevalence is not uniform across regions, but is shaped by context-specific factors, including demographic composition, healthcare access, and regional environments.}

\subsection{\textit{\textbf{Texas county level results}}}

\begin{figure}[tbp]
    \centering
    \includegraphics[width=5.5in]{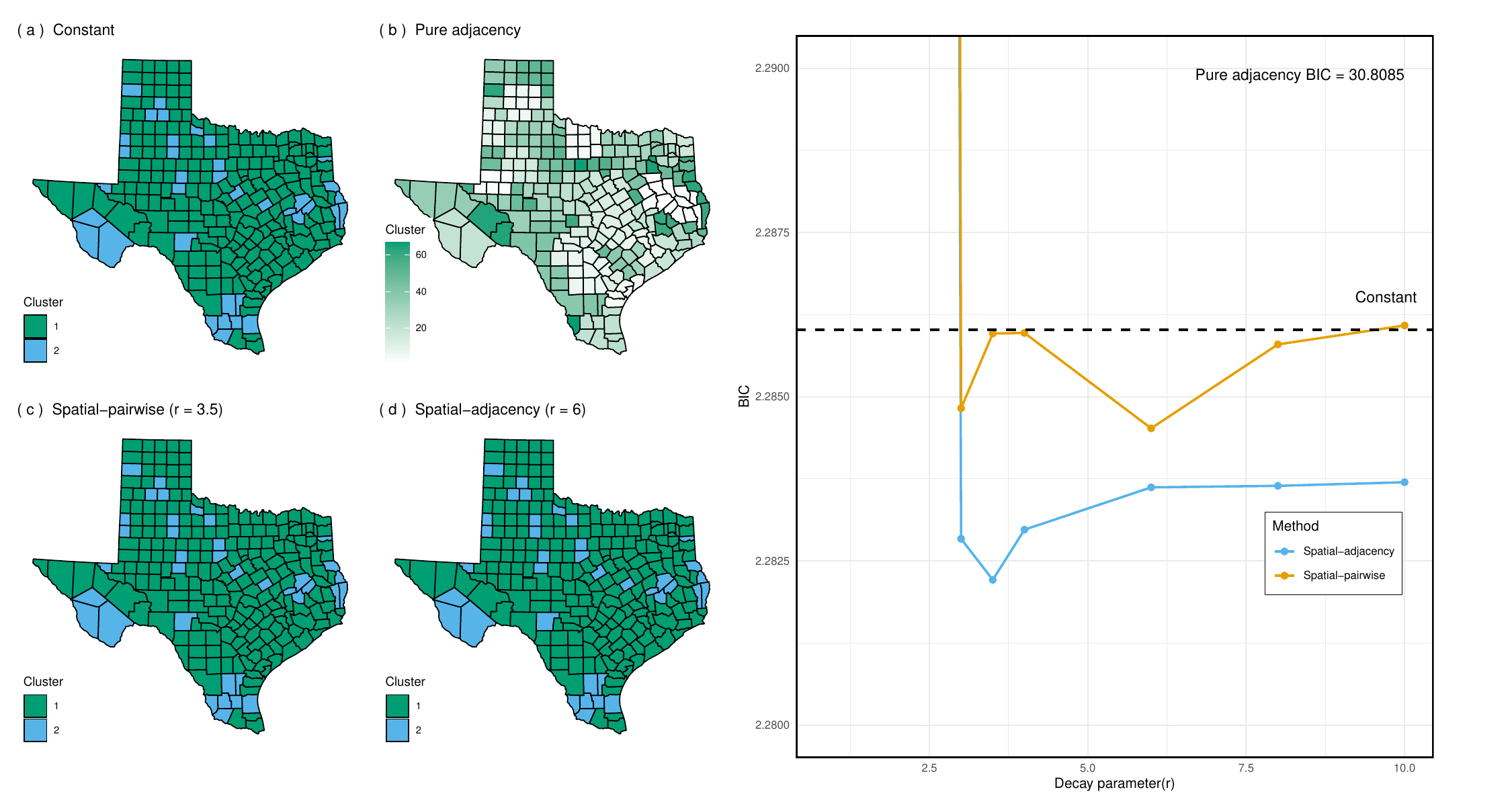}
    \caption{County-level clustering results (left) and BIC comparison (right) under different spatial weighting methods.
    }
    \label{fig:txrealresult}
\end{figure}

{As shown in Figure~\ref{fig:txrealresult}, similar to the state-level analysis, quantitative criteria consistently support the improved performance of the proposed method. The BIC results demonstrate the superiority of the spatial-adjacency method, which consistently achieves lower values across decay parameters compared to other methods, with a BIC = 2.28 at $r = 3.5$.} 

{From a spatial perspective, the pure adjacency method exhibits severe over-clustering, while the proposed approach yields more coherent and interpretable spatial regimes. These clusters likely reflect differences in socioeconomic composition, industrial base, and public health characteristics across regions of Texas, with some border and coastal counties forming one group and much of the central and northern counties forming another.} These findings confirm the target method’s capacity to capture meaningful spatial heterogeneity at a finer geographic resolution, even in the presence of complex county-level variation. 

\begin{table}[htbp]
\centering
\caption{Estimate results of regression coefficients in TX county-level}
\label{tab:tx_beta}
\setlength{\tabcolsep}{3.5pt}
\noindent\makebox[\textwidth]{ % Centering the resized table
\begin{tabular}{p{3cm}rr}
\toprule
& \multicolumn{2}{c}{\textbf{Spatial-adjacency}} \\
& \multicolumn{2}{c}{($\lambda = 1.44$)} \\
& \textbf{Cluster 1} & \textbf{Cluster 2} \\
\midrule
$\bm\beta_{\text{low}}$ & $2.58$ & $0.79$\\
$\bm\beta_{\text{low-med}}$ & $2.07$ & $-2.01$\\
$\bm\beta_{\text{medium}}$ & $-5.85$ & $-0.45$\\
$\bm\beta_{\text{high}}$ & $1.21$ & $1.67$\\
\midrule
$\bm\eta_{\text{Asian}}$ & \multicolumn{2}{c}{$-0.06$}\\
$\bm\eta_{\text{Rural}}$  & \multicolumn{2}{c}{$0.25$}\\
$\bm\eta_{\text{Below 18}}$ & \multicolumn{2}{c}{$2.00$}\\
$\bm\eta_{\text{Children free lunch}}$ & \multicolumn{2}{c}{$0.62$}\\
$\bm\eta_{\text{Adult obesity}}$ & \multicolumn{2}{c}{$-1.39$}\\
$\bm\eta_{\text{Premature mortality}}$ & \multicolumn{2}{c}{$-0.20$}\\
\bottomrule
\end{tabular}
}
\end{table}

{Table~\ref{tab:tx_beta} presents the estimated regression coefficients from the Texas county-level analysis under the spatial-adjacency method. Consistent with the state-level findings, income composition remains an important factor, while the magnitude and direction of effects vary more noticeably across clusters.

In Cluster 1, lower-income components exhibit strong positive associations with COPD prevalence, accompanied by a pronounced negative effect for the medium-income group. In contrast, Cluster 2 shows weaker and less consistent effects among lower-income components, along with a substantially attenuated negative association for the medium-income group. The high-income component is positively associated with the outcome in both clusters, with a stronger effect observed in Cluster 2. 

For non-compositional covariates, several variables, including the proportion of children and participation in free lunch programs, show positive associations, whereas adult obesity and premature mortality display negative relationships. These associations should be interpreted cautiously, as they may reflect regional patterns or differences in data aggregation rather than direct causal effects.

Overall, the county-level results reinforce the presence of spatial heterogeneity and suggest that the relationship between income composition and COPD prevalence becomes more variable at finer geographic scales.
}

%Table \ref{tab:tx_beta} presents the estimated regression coefficients from the Texas county-level analysis using the spatial-adjacency method. The results indicate substantial differences in the effects of income composition across the two identified clusters. Cluster 1 exhibits higher positive effects for the low and low–medium income categories, but a large negative effect for the medium income category, whereas Cluster 2 shows weaker positive or even negative associations for the lower income categories and smaller magnitude negative effects for the medium income group. The high-income category is positively associated with the outcome in both clusters, though the effect is stronger in Cluster 2. For the non-compositional predictors, notable associations include a positive relationship with the proportion of children under 18, and children receiving free lunch, and a negative relationship with adult obesity and premature mortality. These patterns suggest that the spatial-adjacency method successfully distinguishes between county groups with different socioeconomic and demographic profiles, capturing meaningful heterogeneity in both compositional and non-compositional effects.

{Figure~\ref{fig:txrealresult} illustrates the county-level clustering results under different spatial weighting schemes. Similar to state level simulation, clustering is driven by similarities in underlying socioeconomic and health-related characteristics rather than geographic proximity alone. At the county level, where spatial units are more numerous and heterogeneous, such patterns are expected and highlight the complexity of spatial relationships. Rather than corresponding directly to established regional classifications, the detected clusters reflect nuanced and context-dependent structures that may not be fully captured by conventional geographic frameworks.}

\section{Discussion}\label{sec:discussion}

The proposed spatial heterogeneous compositional regression framework with adjacency-based penalization offers a flexible and effective tool for modeling spatially varying effects in the presence of compositional covariates. By incorporating spatial heterogeneity directly into the modeling structure, our approach accommodates localized variations while maintaining a coherent global framework. 
{From a geographic perspective, the proposed method enables the identification of spatial regimes that are not restricted by administrative boundaries or strict geographic proximity. By allowing for non-contiguous clustering, the framework reveals similarities across regions that may share underlying socioeconomic and health-related characteristics, even when they are spatially separated. This provides new insight into spatial inequality and regional heterogeneity that may not be captured by traditional region-based approaches.}
This contribution is particularly valuable in applications where compositional predictors naturally arise—such as epidemiology, environmental science, and economics—since it enables researchers to uncover region-specific associations without discarding important global patterns. Moreover, adjacency-based penalization provides a principled way to borrow strength across neighboring regions, improving estimation accuracy and interpretability while respecting the underlying spatial structure. Taken together, our contribution lies in bridging the methodological gap between compositional data analysis and spatially heterogeneous regression, offering a framework that is theoretically grounded and practically useful.

Looking ahead, several promising directions for future work emerge. One avenue is to extend the framework to handle multivariate or hierarchically structured compositional data, where dependencies may exist both within and across multiple compositional systems. Such extensions would broaden the applicability of the method to complex real-world data sets, such as multi-level biological measurements or socioeconomic indicators collected at nested spatial scales. Another important direction involves developing scalable algorithms to accommodate high-dimensional compositional covariates while maintaining computational efficiency. Finally, future research could focus on establishing theoretical guarantees, such as consistency and convergence rates, under more general dependence structures. These efforts would not only strengthen the theoretical foundation of the framework but also enhance its robustness and adaptability across a wide range of scientific applications. {These extensions would further enhance the applicability of the framework in spatial and geographic research contexts.}

\section*{Funding}
This work was supported by National Science Foundation under Grant SES-2412922.
%[removed for peer review]

\section*{Disclosure Statement}
The authors report there are no competing financial or non-financial interests to declare.

\section*{Data availability statement}
All data analyzed in this study are publicly available from the Centers for Disease Control and Prevention (CDC), the U.S. Census Bureau, the U.S. Bureau of Economic Analysis, and County Health Rankings \& Roadmaps. URLs to each dataset are provided in the Reference. No proprietary or confidential data were used.
%[removed for peer review]

\bibliographystyle{apacite}

\bibliography{Bib}

@misc{dshs_regions,
  author = {{Texas Department of State Health Services}},
  title = {Public Health Regions},
  year = {2023},
  url = {https://www.dshs.texas.gov/regions/},
  note = {Accessed: 2026-04-10}
}

@article{Pleasants2016,
  author  = {Pleasants, R. A. and Riley, I. L. and Mannino, D. M.},
  title   = {Defining and targeting health disparities in chronic obstructive pulmonary disease},
  journal = {Int J Chron Obstruct Pulmon Dis},
  year    = {2016},
  volume  = {11},
  pages   = {2475--2496},
  doi     = {10.2147/COPD.S79077}
}

@article{grigsby2016socioeconomic,
  title={Socioeconomic status and {COPD} among low- and middle-income countries},
  author={Grigsby, M. and Siddharthan, T. and Chowdhury, M. A. and Siddiquee, A. and Rubinstein, A. and Sobrino, E. and Miranda, J. J. and Bernabe-Ortiz, A. and Alam, D. and Checkley, W.},
  journal={International Journal of Chronic Obstructive Pulmonary Disease},
  volume={11},
  pages={2497--2507},
  year={2016},
  month={Oct},
  publisher={Dove Press},
  doi={10.2147/COPD.S111145},
  pmid={27785006},
  pmcid={PMC5065097}
}

@article{anselin2024EndogenousSpatial,
  title = {Endogenous Spatial Regimes},
  author = {Anselin, Luc and Amaral, Pedro},
  year = {2024},
  journal = {Journal of geographical systems},
  volume = {26},
  number = {2},
  pages = {209--234},
  publisher = {Springer Berlin Heidelberg},
  address = {Berlin/Heidelberg},
  issn = {1435-5930},
  doi = {10.1007/s10109-023-00411-2},
  langid = {english}
}

@article{li2019spatial,
  title={Spatial homogeneity pursuit of regression coefficients for large datasets},
  author={Li, Furong and Sang, Huiyan},
  journal={Journal of the American Statistical Association},
  year={2019},
  publisher={Taylor \& Francis}
}

@article{wang2024scanner,
  title={Scanner: Simultaneously temporal trend and spatial cluster detection for spatial-temporal data},
  author={Wang, Xin and Zhang, Xin},
  journal={Environmetrics},
  volume={35},
  number={5},
  pages={e2849},
  year={2024},
  publisher={Wiley Online Library}
}

@article{lin2014variable,
  title={Variable selection in regression with compositional covariates},
  author={Lin, Wei and Shi, Pixu and Feng, Rui and Li, Hongzhe},
  journal={Biometrika},
  volume={101},
  number={4},
  pages={785--797},
  year={2014},
  publisher={Oxford University Press}
}

@article{kamenetsky2022regularized,
  title={Regularized spatial and spatio-temporal cluster detection},
  author={Kamenetsky, Maria E and Lee, Junho and Zhu, Jun and Gangnon, Ronald E},
  journal={Spatial and Spatio-temporal Epidemiology},
  volume={41},
  pages={100462},
  year={2022},
  publisher={Elsevier}
}

@article{ma_exploration_2020,
	title = {Exploration of {Heterogeneous} {Treatment} {Effects} via {Concave} {Fusion}},
	volume = {16},
	issn = {1557-4679, 2194-573X},
	url = {https://www.degruyter.com/document/doi/10.1515/ijb-2018-0026/html},
	doi = {10.1515/ijb-2018-0026},
	language = {en},
	number = {1},
	urldate = {2023-09-29},
	journal = {The International Journal of Biostatistics},
	author = {Ma, Shujie and Huang, Jian and Zhang, Zhiwei and Liu, Mingming},
	month = may,
	year = {2020},
	pages = {20180026},
}

@article{aitchison_statistical_1982,
	title = {The {Statistical} {Analysis} of {Compositional} {Data}},
	volume = {44},
	issn = {0035-9246},
	url = {https://doi.org/10.1111/j.2517-6161.1982.tb01195.x},
	doi = {10.1111/j.2517-6161.1982.tb01195.x},
	number = {2},
	urldate = {2024-11-26},
	journal = {Journal of the Royal Statistical Society: Series B (Methodological)},
	author = {Aitchison, J.},
	month = jan,
	year = {1982},
	pages = {139--160},
}

@article{boyd_distributed_2010,
	title = {Distributed {Optimization} and {Statistical} {Learning} via the {Alternating} {Direction} {Method} of {Multipliers}},
	volume = {3},
	issn = {1935-8237, 1935-8245},
	url = {http://www.nowpublishers.com/article/Details/MAL-016},
	doi = {10.1561/2200000016},
	language = {en},
	number = {1},
	urldate = {2024-11-26},
	journal = {Foundations and Trends® in Machine Learning},
	author = {Boyd, Stephen},
	year = {2010},
	pages = {1--122},
}

@article{brunsdon_geographically_1996,
	title = {Geographically {Weighted} {Regression}: {A} {Method} for {Exploring} {Spatial} {Nonstationarity}},
	volume = {28},
	copyright = {1996 The Ohio State University},
	issn = {1538-4632},
	shorttitle = {Geographically {Weighted} {Regression}},
	url = {https://onlinelibrary.wiley.com/doi/abs/10.1111/j.1538-4632.1996.tb00936.x},
	doi = {10.1111/j.1538-4632.1996.tb00936.x},
	language = {en},
	number = {4},
	urldate = {2024-11-26},
	journal = {Geographical Analysis},
	author = {Brunsdon, Chris and Fotheringham, A. Stewart and Charlton, Martin E.},
	year = {1996},
	note = {\_eprint: https://onlinelibrary.wiley.com/doi/pdf/10.1111/j.1538-4632.1996.tb00936.x},
	pages = {281--298},
}

@article{gelfand_spatial_2003,
	title = {Spatial {Modeling} {With} {Spatially} {Varying} {Coefficient} {Processes}},
	volume = {98},
	issn = {0162-1459},
	doi = {10.1198/016214503000170},
	language = {eng},
	number = {462},
	journal = {Journal of the American Statistical Association},
	author = {Gelfand, Alan E. and Kim, Hyon-Jung and Sirmans, C. F. and Banerjee, Sudipto},
	year = {2003},
	pmid = {39421645},
	pmcid = {PMC11484471},
	pages = {387--396},
}

@article{greenacre2021compositional,
  title={Compositional data analysis},
  author={Greenacre, Michael},
  journal={Annual Review of Statistics and its Application},
  volume={8},
  number={1},
  pages={271--299},
  year={2021},
  publisher={Annual Reviews}
}

@incollection{macqueen_methods_1967,
	title = {Some methods for classification and analysis of multivariate observations},
	volume = {5.1},
	url = {https://projecteuclid.org/ebooks/berkeley-symposium-on-mathematical-statistics-and-probability/Proceedings-of-the-Fifth-Berkeley-Symposium-on-Mathematical-Statistics-and/chapter/Some-methods-for-classification-and-analysis-of-multivariate-observations/bsmsp/1200512992},
	urldate = {2025-01-15},
	booktitle = {Proceedings of the {Fifth} {Berkeley} {Symposium} on {Mathematical} {Statistics} and {Probability}, {Volume} 1: {Statistics}},
	publisher = {University of California Press},
	author = {MacQueen, J.},
	month = jan,
	year = {1967},
	pages = {281--298},
}

@article{tibshirani_regression_1996,
	title = {Regression {Shrinkage} and {Selection} {Via} the {Lasso}},
	volume = {58},
	copyright = {https://academic.oup.com/journals/pages/open\_access/funder\_policies/chorus/standard\_publication\_model},
	issn = {1369-7412, 1467-9868},
	url = {https://academic.oup.com/jrsssb/article/58/1/267/7027929},
	doi = {10.1111/j.2517-6161.1996.tb02080.x},
	language = {en},
	number = {1},
	urldate = {2024-11-30},
	journal = {Journal of the Royal Statistical Society Series B: Statistical Methodology},
	author = {Tibshirani, Robert},
	month = jan,
	year = {1996},
	pages = {267--288},
}

@misc{chr_2022,
  author       = {{County Health Rankings \& Roadmaps}},
  title        = {{2022 County Health Rankings National Data (Analytic Data 2022) [Data set]}},
  howpublished = {University of Wisconsin Population Health Institute},
  year         = {2024},
  url          = {https://www.countyhealthrankings.org/explore-health-rankings/rankings-data-documentation},
  note         = {Direct download link: \url{https://www.countyhealthrankings.org/sites/default/files/media/document/analytic_data2022.csv}}
}

@misc{bea_gdp_state_2024,
  author       = {{U.S. Bureau of Economic Analysis}},
  title        = {GDP by State},
  howpublished = {U.S. Bureau of Economic Analysis (BEA) website},
  year         = {2024},
  url          = {https://www.bea.gov/data/gdp/gdp-state},
  urldate      = {2024-12-05}, 
  note         = {Accessed December 5, 2024}
}

@misc{us_census_acs_2024,
  author       = {{U.S. Census Bureau}},
  title        = {{American Community Survey (ACS)}},
  howpublished = {U.S. Census Bureau website},
  year         = {2024},
  url          = {https://www.census.gov/programs-surveys/acs},
  urldate      = {2025-01-15},
  note         = {Accessed 2024. The American Community Survey is the premier source for information about America's changing population, housing, and workforce.}
}

@misc{cdc_brfss_2024,
  author       = {{BRFSS}},
  title        = {{Behavioral Risk Factor Surveillance System (BRFSS) Prevalence Data}},
  howpublished = {CDC Website},
  year         = {2024},
  url          = {https://www.cdc.gov/brfss/brfssprevalence/index.html},
  note         = {[Data set] Accessed in 2024}
}

@misc{cdc_places_2024,
  author       = {{PLACES}},
  title        = {{PLACES: Local Data for Better Health}},
  howpublished = {CDC Website},
  year         = {2024},
  url          = {https://www.cdc.gov/places/index.html},
  note         = {[Data set] Accessed in 2024}
}

@article{egozcue_isometric_2003,
	title = {Isometric {Logratio} {Transformations} for {Compositional} {Data} {Analysis}},
	language = {en},
	journal = {Mathematical Geology},
	author = {Egozcue, J J and Pawlowsky-Glahn, V and Mateu-Figueras, G and Barcelo-Vidal, C},
	year = {2003},
}

@misc{zhang_nearly_2010,
	title = {Nearly unbiased variable selection under minimax concave penalty},
	url = {http://arxiv.org/abs/1002.4734},
	doi = {10.48550/arXiv.1002.4734},
	urldate = {2024-12-06},
	publisher = {arXiv},
	author = {Zhang, Cun-Hui},
	month = feb,
	year = {2010},
	note = {arXiv:1002.4734},
}

@article{wang_tuning_2007,
	title = {Tuning parameter selectors for the smoothly clipped absolute deviation method},
	volume = {94},
	issn = {0006-3444},
	url = {https://www.ncbi.nlm.nih.gov/pmc/articles/PMC2663963/},
	doi = {10.1093/biomet/asm053},
	number = {3},
	urldate = {2024-12-06},
	journal = {Biometrika},
	author = {Wang, Hansheng and Li, Runze and Tsai, Chih-Ling},
	month = aug,
	year = {2007},
	pmid = {19343105},
	pmcid = {PMC2663963},
	pages = {553--568},
}

@article{rand_objective_1971,
	title = {Objective {Criteria} for the {Evaluation} of {Clustering} {Methods}},
	volume = {66},
	issn = {0162-1459},
	doi = {10.1080/01621459.1971.10482356},
	language = {eng},
	number = {336},
	journal = {Journal of the American Statistical Association},
	author = {Rand, William M.},
	year = {1971},
	note = {Publisher: Taylor \& Francis Group},
	pages = {846--850},
}

@article{lee_cluster_2017,
	title = {Cluster detection of spatial regression coefficients},
	volume = {36},
	copyright = {http://onlinelibrary.wiley.com/termsAndConditions\#vor},
	issn = {0277-6715, 1097-0258},
	url = {https://onlinelibrary.wiley.com/doi/10.1002/sim.7172},
	doi = {10.1002/sim.7172},
	language = {en},
	number = {7},
	urldate = {2024-12-10},
	journal = {Statistics in Medicine},
	author = {Lee, Junho and Gangnon, Ronald E. and Zhu, Jun},
	month = mar,
	year = {2017},
	pages = {1118--1133},
}

@article{suarez_bayesian_2016,
	title = {Bayesian {Clustering} of {Functional} {Data} {Using} {Local} {Features}},
	volume = {11},
	issn = {1936-0975},
	url = {https://projecteuclid.org/journals/bayesian-analysis/volume-11/issue-1/Bayesian-Clustering-of-Functional-Data-Using-Local-Features/10.1214/14-BA925.full},
	doi = {10.1214/14-BA925},
	language = {en},
	number = {1},
	urldate = {2024-12-10},
	journal = {Bayesian Analysis},
	author = {Suarez, Adam Justin and Ghosal, Subhashis},
	month = mar,
	year = {2016},
}

@article{jiang_clustering_2012,
	title = {Clustering {Random} {Curves} {Under} {Spatial} {Interdependence} {With} {Application} to {Service} {Accessibility}},
	volume = {54},
	issn = {0040-1706, 1537-2723},
	url = {http://www.tandfonline.com/doi/abs/10.1080/00401706.2012.657106},
	doi = {10.1080/00401706.2012.657106},
	language = {en},
	number = {2},
	urldate = {2024-12-10},
	journal = {Technometrics},
	author = {Jiang, Huijing and Serban, Nicoleta},
	month = may,
	year = {2012},
	pages = {108--119},
}

@article{giraldo_hierarchical_2012,
	title = {Hierarchical clustering of spatially correlated functional data},
	volume = {66},
	issn = {0039-0402, 1467-9574},
	url = {https://onlinelibrary.wiley.com/doi/10.1111/j.1467-9574.2012.00522.x},
	doi = {10.1111/j.1467-9574.2012.00522.x},
	language = {en},
	number = {4},
	urldate = {2024-12-10},
	journal = {Statistica Neerlandica},
	author = {Giraldo, R. and Delicado, P. and Mateu, J.},
	month = nov,
	year = {2012},
	pages = {403--421},
}

@book{everitt_finite_1981,
	address = {Dordrecht},
	title = {Finite {Mixture} {Distributions}},
	copyright = {http://www.springer.com/tdm},
	isbn = {978-94-009-5899-9 978-94-009-5897-5},
	url = {http://link.springer.com/10.1007/978-94-009-5897-5},
	language = {en},
	urldate = {2025-01-15},
	publisher = {Springer Netherlands},
	author = {Everitt, B. S. and Hand, D. J.},
	year = {1981},
	doi = {10.1007/978-94-009-5897-5},
}

@article{mcnicholas_model-based_2010,
	title = {Model-based classification using latent {Gaussian} mixture models},
	volume = {140},
	issn = {0378-3758},
	url = {https://www.sciencedirect.com/science/article/pii/S0378375809003607},
	doi = {10.1016/j.jspi.2009.11.006},
	number = {5},
	urldate = {2024-12-29},
	journal = {Journal of Statistical Planning and Inference},
	author = {McNicholas, Paul D.},
	month = may,
	year = {2010},
	pages = {1175--1181},
}

@article{banfield_model-based_1993,
	title = {Model-{Based} {Gaussian} and {Non}-{Gaussian} {Clustering}},
	volume = {49},
	issn = {0006-341X},
	url = {https://www.jstor.org/stable/2532201},
	doi = {10.2307/2532201},
	number = {3},
	urldate = {2024-12-29},
	journal = {Biometrics},
	author = {Banfield, Jeffrey D. and Raftery, Adrian E.},
	year = {1993},
	note = {Publisher: International Biometric Society},
	pages = {803--821},
}

@article{wei_latent_2013,
	title = {Latent {Supervised} {Learning}},
	volume = {108},
	issn = {0162-1459},
	url = {https://www.ncbi.nlm.nih.gov/pmc/articles/PMC3848255/},
	doi = {10.1080/01621459.2013.789695},
	number = {503},
	urldate = {2024-12-29},
	journal = {Journal of the American Statistical Association},
	author = {Wei, Susan and Kosorok, Michael R.},
	month = jul,
	year = {2013},
	pmid = {24319303},
	pmcid = {PMC3848255},
	pages = {10.1080/01621459.2013.789695},
}

@article{shen_inference_2015,
	title = {Inference for {Subgroup} {Analysis} {With} a {Structured} {Logistic}-{Normal} {Mixture} {Model}},
	volume = {110},
	issn = {0162-1459},
	url = {https://www.jstor.org/stable/24739305},
	number = {509},
	urldate = {2024-12-29},
	journal = {Journal of the American Statistical Association},
	author = {Shen, Juan and He, Xuming},
	year = {2015},
	note = {Publisher: [American Statistical Association, Taylor \& Francis, Ltd.]},
	pages = {303--312},
}

@misc{pew2024middleclass,
  author = {{Pew Research Center}},
  title = {The State of the American Middle Class},
  year = {2024},
  note = {Published May 31, 2024},
  url = {https://www.pewresearch.org/race-and-ethnicity/2024/05/31/the-state-of-the-american-middle-class/}
}

@Article{Hornik2005clue,
  author    = {Kurt Hornik},
  title     = {A CLUE for CLUster Ensembles},
  journal   = {Journal of Statistical Software},
  year      = {2005},
  volume    = {14},
  number    = {12},
  doi       = {10.18637/jss.v014.i12},
}

@article{prates_transformed_2015,
	title = {Transformed {Gaussian} {Markov} {Random} {Fields} and {Spatial} {Modeling}},
	volume = {14},
	issn = {22116753},
	url = {http://arxiv.org/abs/1205.5467},
	doi = {10.1016/j.spasta.2015.07.004},
	urldate = {2024-12-29},
	journal = {Spatial Statistics},
	author = {Prates, Marcos O. and Dey, Dipak K. and Willig, Michael R. and Yan, Jun},
	month = nov,
	year = {2015},
	note = {arXiv:1205.5467 [stat]},
	pages = {382--399},
}

@article{fan_variable_2001,
	title = {Variable {Selection} via {Nonconcave} {Penalized} {Likelihood} and {Its} {Oracle} {Properties}},
	volume = {96},
	issn = {0162-1459},
	url = {https://www.jstor.org/stable/3085904},
	number = {456},
	urldate = {2025-01-15},
	journal = {Journal of the American Statistical Association},
	author = {Fan, Jianqing and Li, Runze},
	year = {2001},
	note = {Publisher: [American Statistical Association, Taylor \& Francis, Ltd.]},
	pages = {1348--1360},
}

@article{Tobler_1970_computer,
 ISSN = {00130095, 19448287},
 URL = {http://www.jstor.org/stable/143141},
 author = {W. R. Tobler},
 journal = {Economic Geography},
 pages = {234--240},
 publisher = {[Clark University, Wiley]},
 title = {A Computer Movie Simulating Urban Growth in the Detroit Region},
 urldate = {2025-01-20},
 volume = {46},
 year = {1970}
}

@article{Subramanian_2004_ineq,
    author = {Subramanian, S. V. and Kawachi, Ichiro},
    title = {Income Inequality and Health: What Have We Learned So Far?},
    journal = {Epidemiologic Reviews},
    volume = {26},
    number = {1},
    pages = {78-91},
    year = {2004},
    month = {07},
    issn = {0193-936X},
    doi = {10.1093/epirev/mxh003},
    url = {https://doi.org/10.1093/epirev/mxh003},
    eprint = {https://academic.oup.com/epirev/article-pdf/26/1/78/1087997/mxh003.pdf},
}

@article{quin_2020_income,
  title={Income inequalities in the risk of potentially avoidable hospitalisation for chronic obstructive pulmonary disease: a population data linkage analysis},
  author={Quinn, Nicholas and Gupta, Neeru},
  journal={International Journal of Population Data Science},
  volume={5},
  number={1},
  year={2020},
  publisher={Swansea University},
  doi={10.23889/ijpds.v5i1.1388}
}

@book{snyder1987map,
  author = {Snyder, John P.},
  title = {Map Projections--A Working Manual},
  series = {U.S. Geological Survey Professional Paper},
  volume = {1395},
  year = {1987},
  publisher = {U.S. Government Printing Office}
}

@book{duda2001pattern,
  author = {Duda, Richard O. and Hart, Peter E. and Stork, David G.},
  title = {Pattern Classification},
  edition = {2nd},
  year = {2001},
  publisher = {John Wiley \& Sons}
}

@article{geng2022bayesian,
  title={Bayesian spatial homogeneity pursuit for survival data with an application to the SEER respiratory cancer data},
  author={Geng, Lijiang and Hu, Guanyu},
  journal={Biometrics},
  volume={78},
  number={2},
  pages={536--547},
  year={2022},
  publisher={Wiley Online Library}
}

@article{xue2020geographically,
  title={Geographically weighted Cox regression for prostate cancer survival data in Louisiana},
  author={Xue, Yishu and Schifano, Elizabeth D and Hu, Guanyu},
  journal={Geographical Analysis},
  volume={52},
  number={4},
  pages={570--587},
  year={2020},
  publisher={Wiley Online Library}
}

\appendix

\externaldocument{main}

%other

\doublespacing
% \addtolength{\oddsidemargin}{-.5in}%
% \addtolength{\evensidemargin}{-.5in}%
% \addtolength{\textwidth}{1in}%
% \addtolength{\textheight}{-.3in}%
%\addtolength{\topmargin}{-.8in}%
\def\spacingset#1{\renewcommand{\baselinestretch}%
{#1}\small\normalsize} \spacingset{1.5}
\def\blue{\color{blue}}
\def\red{\color{red}}
\def\mR{\mathbb{R}}
\def\M{\bm{M}}
\def\Q{\bm{Q}}
\def\X{\bm{X}}
\def\defeq{\stackrel{\mathrm{def}}{=}}  % for definitions
\def\balpha{\bfsym \alpha}
\def\bbeta{\bfsym \beta}
\def\bgamma{\bfsym \gamma}             \def\bGamma{\bfsym \Gamma}
\def\bdelta{\bfsym {\delta}}           \def\bDelta {\bfsym {\Delta}}
\def\bfeta{\bfsym {\eta}}              \def\bfEta {\bfsym {\Eta}}
\def\bmu{\bfsym {\mu}}                 \def\bMu {\bfsym {\Mu}}
\def\bnu{\bfsym {\nu}}
\def\btheta{\bfsym {\theta}}           \def\bTheta {\bfsym {\Theta}}
\def\beps{\bfsym \varepsilon}          \def\bepsilon{\bfsym \varepsilon}
\def\bsigma{\bfsym \sigma}             \def\bSigma{\bfsym \Sigma}
\def\blambda {\bfsym {\lambda}}        \def\bLambda {\bfsym {\Lambda}}
\def\bomega {\bfsym {\omega}}          \def\bOmega {\bfsym {\Omega}}
\def\brho   {\bfsym {\rho}}
\def\btau{\bfsym {\tau}}
\def\bxi{\bfsym {\xi}}          \def\bXi{\bfsym {\Xi}}
\def\bzeta{\bfsym {\zeta}}
\def\mY{\mathbb{Y}}
\def\mZ{\mathbb{Z}}
\def\mX{\mathbb{X}}
\def\mR{\mathbb{R}}
\def\bX{\mathbf{X}}
\def\zero{\mathbf{0}}
\def\mG{\mathcal{G}}

\def\red{\color{red}}
\def\blue{\color{blue}}

%\title{\Large\bfseries Supplementary for Linking COPD Prevalence with Income Distribution: A Spatial Heterogeneous Compositional Regression via Geographically Weighted Penalized Approach}
%\author{}
%\date{}

%\input{./notation_revised}

%\maketitle

\section{Estimation procedure via Alternating Direction Method of Multipliers (ADMM) algorithm}
\label{sec:Appendix_ADMM}

In this section, we discuss the computational algorithm using ADMM, a summary of this process is summarized in Main Algorithm 1. 

Let $\bm\delta_{ij}=\bbeta_i-\bbeta_j$ for $1\leq i<j\leq n$.
Main Eq.(2) is equivalent to 

\begin{align}
    L_0(\bm{\beta}, \bm{\eta}, \bm{\alpha}) &= \frac{1}{2} \sum_{i=1}^n \left( y_i - \bm{x}_{1i}^\top \bm{\beta}_i - \bm{x}_{2i}^\top \bm{\eta}_i \right)^2 + \sum_{1 \leq i < j \leq n} p_\gamma \left( \|\bm{\delta}_{ij}\|, \lambda, \omega \right), \nonumber \\[1em]
    & \bm\beta_i-\bm\beta_j = \bm{\delta}_{ij}, \quad \forall 1 \leq i < j \leq n,
\label{eq:miniquestion}
\end{align}
with $\bm\alpha = \left\{ \balpha_{ij}^\top, i<j \right\} ^\top$ as the Lagrange multipliers, $\vartheta > 0$ as the penalty parameter, which is a hyperparameter in ADMM algorithm, the augmented Lagrangian function for \eqref{eq:miniquestion} is

\begin{align}
L(\bm{\beta}, \bm{\eta}, \bm{\delta}, \bm{\alpha}) = L_0(\bm{\beta}, \bm{\eta}, \bm{\delta}) + \sum_{1 \leq i < j \leq n} \langle \bm{\alpha}_{ij}, \bm\beta_i-\bm\beta_j - \bm{\delta}_{ij} \rangle + \frac{\vartheta}{2} \sum_{1 \leq i < j \leq n} \| \bm\beta_i-\bm\beta_j - \bm\delta_{ij} \|^2,
\end{align}

where $\bdelta = \left\{ \bdelta_{ij}^\top, i<j \right\} ^\top$, $\langle a, b \rangle = a ^\top b$ is inner product with $a$ and $b$ having same dimension.

The optimization problem is solved iteratively by updating the primal variables $(\bm{\beta}, \bm{\eta}, \bm{\delta})$ and dual variables $\bm{\alpha}$ using ADMM algorithm. 

At $m$-th iteration of ADMM, the minimization of parameters are: 

\begin{align}
(\bm\beta^{m+1}, \bm\eta^{m+1}) &= \arg\min_{\bm\beta, \bm\eta} L(\bm\beta, \bm\eta, \bm\delta^m, \bm\alpha^m), 
\label{eq:mini_beta_eta}
\end{align}

\begin{align*}
\bm\delta^{m+1} &= \arg \min_{\bm\delta} L(\bm\beta^{m+1}, \bm\eta^{m+1}, \bm\delta, \bm\alpha^m),
%\label{eq:mini_delta}
\end{align*}

Dual Update: 
\begin{align}
\bm\alpha_{ij}^{m+1} &= \bm\alpha_{ij}^m + \vartheta (\bm\beta_i^{m+1} - \bm\beta_j^{m+1} - \bm\delta_{ij}^{m+1}).
\label{eq:dual}
\end{align}

To solve the optimization problem iteratively, we focus initially on the subproblem involving the variables $\bm{\beta}$ and $\bm{\eta}$, as outlined in \eqref{eq:mini_beta_eta}. The augmented Lagrangian method separates the variables into independent updates, allowing for the decomposition of the problem into manageable subproblems. At each iteration $m$, the current values of $\bm{\delta}^m$ and $\bm{\alpha}^m$ are treated as fixed. The objective is to minimize the augmented Lagrangian with respect to $\bm{\beta}$ and $\bm{\eta}$

This minimization task can be reformulated as a least-squares problem by carefully expanding and rearranging the terms, as demonstrated in the following equation:

\begin{align*}
f(\bm\beta, \bm\eta) = \frac{1}{2} \sum_{i=1}^n \left( y_i - \bm{x}_{1i}^\top \bm\beta_i - \bm{x}_{2i}^\top \bm\eta_i \right)^2 + \frac{\vartheta}{2} \sum_{i < j} \|\bm\beta_i - \bm\beta_j - \bm\delta_{ij}^m + \vartheta^{-1} \bm\alpha_{ij}^m\|^2 + C, 
\end{align*}
with $e_i$ is $n \times 1$ vector whose $i$-th element is $1$ and the remaining ones are $0$. Set $D = \{(e_i - e_j), i < j\}^\top$. Thus $D$ is of size $\frac{n(n-1)}{2} \times n $, where $n$ is the dimension of the vectors $e_i$, each row of $D$ corresponds to a pair $(i, j)$ with $i < j$, representing the difference between the standard basis vectors $e_i$ and $e_j$. Specifically, for each pair $(i, j)$, The $i$-th column of the row is $+1$, The $j$-th column of the row is $-1$, and all other columns are $0$.

Set $\mathbf{A} = D \otimes \mathbf{I_p}$. With $\mathbf{I_p}$ is the $p \times p$ identity matrix,  $\otimes$ denotes the Kronecker product, $\mathbf{A}$ has a size of $\frac{n(n-1)}{2} \cdot p \times n \cdot p$, each block in $\mathbf{A}$ represents the pairwise differences of parameters, with each difference replicated across $p$ -dimensional identity matrices.

$C$ is a constant independent of $(\bm\beta, \bm\eta)$, by collecting the constant parts into $C$, we rewrite  $f(\bm\beta, \bm\eta)$ in a more compact form as:

\begin{align}
f(\bm\beta, \bm\eta) = \frac{1}{2} \|\mathbf{X_1}\bm\beta + \mathbf{X_2}\bm\eta - y\|^2 + \frac{\vartheta}{2} \|\mathbf{A}\bm\beta - \bm\delta^m + \vartheta^{-1} \bm\alpha^m\|^2 + C.
\label{eq:compact form}
\end{align}

Set $\mathbf{Q} = \mathbf{I_n} - \mathbf{X_2}(\mathbf{X_2}^\top \mathbf{X_2})^{-1} \mathbf{X_2}^\top$. Thus, at $m$-th step, for given $\delta^m$ and $\alpha^m$ are treated as fixed variable, the $\bm\beta_{ij}$ is updated using following equation: 

\begin{align}
\bm\beta^{m+1} &= (\mathbf{X}_1^\top \mathbf{Q} \mathbf{X}_1 + \vartheta \mathbf{A}^\top \mathbf{A})^{-1} \left[ \mathbf{X}_1^\top \mathbf{Q} y + \vartheta \mathbf{A}^\top (\bm\delta^m - \vartheta^{-1} \bm\alpha^m) \right].
\label{eq:beta_update}
\end{align}

In the \eqref{eq:beta_update} we have
\begin{align*}
\mathbf{A}^\top (\bm\delta^m - \vartheta^{-1} \bm\alpha^m) = (D^\top \otimes \mathbf{I_p}) \text{vec}((\bm\Delta^m - \vartheta^{-1} \bm{V}^m) D),
\end{align*}
where $\bm\Delta^m = \{\bm\delta_{ij}^m, i < j\}_{p \times n(n-1)/2}$, $\bm{V}^m = \{\bm\alpha_{ij}^m, i < j\}_{p \times n(n-1)/2}$, and we can rewrite $\beta^{m+1}$ as: 

\begin{align*}
\bm\beta^{m+1} = (\mathbf{X_1}^\top \mathbf{Q} \mathbf{X_1} + \vartheta \mathbf{A}^\top \mathbf{A})^{-1} \left[ \mathbf{X_1}^\top \mathbf{Q} y + \vartheta \text{vec}((\bm\Delta^m - \vartheta^{-1} \mathbf{V}^m) D) \right].
\end{align*}

The $\bm\eta_{ij}$ is then updated using following equation: 

\begin{align}
\bm\eta^{m+1} &= (\mathbf{X_2}^\top \mathbf{X_2})^{-1} \mathbf{X_2}^\top \left( y - \mathbf{X_1}\bm\beta^{m+1} \right).
\label{eq:eta_update}
\end{align}

After collecting the constant parts, the minimization question to find $\bm\delta^{m+1}$ is equal to: 

\begin{align}
\bm\delta^{m+1} = \arg \min_{\bm\delta} \frac{\vartheta}{2} \| \bm\beta_i^m - \bm\beta_j^m + \vartheta^{-1} \bm\alpha_{ij}^m - \bm\delta_{ij} \|^2 + p_\gamma(\|\bm\delta_{ij}\|, \lambda).
\label{eq:d_update}
\end{align}

Applying MCP with $\gamma > 1/\vartheta$, with $\bm\zeta_{ij}^m = \bm\beta_i^m - \bm\beta_j^m + \vartheta^{-1} \bm\alpha_{ij}^m$, the updated $\bm\delta_{ij}$ would be: 

\begin{align*}
\bm\delta_{ij}^{m+1} =
\begin{cases} 
\frac{S(\bm\zeta_{ij}^m, \lambda / \vartheta)}{1- 1/\gamma\vartheta}, & \text{if } \|\bm\zeta_{ij}^m\| \leq \gamma (\lambda \omega), \\[1em] 
\bm\zeta_{ij}^m, & \text{if } \|\bm\zeta_{ij}^m\| > \gamma (\lambda \omega),
\end{cases}
\end{align*}
with 
\begin{align*}
S(\bm{z},t) = (1 - t / \|\bm{z}\|)_+ \bm{z}), 
\end{align*}
and 

\begin{align*}
(x)_+ = 
\begin{cases} 
x & \text{if } x>0, \\[1em] 
0& \text{Otherwise}.
\end{cases}
\end{align*}

Finally, operate the dual update step for $\bm\alpha_{ij}$ using \eqref{eq:dual}. 

\section{Hyper-parameter selection and parameter initialization}
\label{subsec:Initialization}

To ensure efficient convergence and enhance the numerical stability of the ADMM algorithm, we carefully initialize the parameters $\bm{\beta}$, $\bm{\eta}$, $\bm\delta$, $\lambda$, and $\bm\alpha$.

The regularization parameter $\lambda$, as the tuning parameter, governs the degree of sparsity and is initialized by exploring a range of possible values. The initial value $\lambda^{(0)}$ is chosen to be very small ($\lambda^{(0)} = 0.001$) such that all pairwise differences $\|\bm{\beta}_i - \bm{\beta}_j\|$ are close to zero to ensure the starting point emphasizes a high level of regularization. This ensures that the starting point emphasizes a high level of regularization, to ensure stability, prevent overfitting, and guide the optimization process efficiently.

To ensure that the optimization process starts from a stable and meaningful point, the ridge fusion criterion is used: 

\begin{align*}
L_R(\bm\beta, \bm\eta) = \frac{1}{2} \| \mathbf{X_1}\bm\beta + \mathbf{X_2}\bm\eta - y \|^2 + \frac{\lambda^{(0)}}{2} \sum_{1 \leq i < j \leq n} \|\beta_i - \beta_j\|^2. 
\end{align*}

Reuse the $\mathbf{A}$ defined in \eqref{eq:compact form}, we can rewrite the above question for a clear look as
\begin{align*}
L_R(\bm\beta, \bm\eta) = \frac{1}{2} \| \mathbf{X_1}\bm\beta + \mathbf{X_2}\bm\eta - y  \|^2 + \frac{\lambda^*}{2} \|\mathbf{A}\beta\|^2.
\end{align*}

Utilize $\bm{Q}$ defined in \eqref{eq:beta_update}, the initial value of $\bm{\beta}$ and $\bm{\eta}$ can be calculated by: 

\begin{align*}
\bm\beta^{(0)}(\lambda^{(0)}) = (\mathbf{X_1}^\top \mathbf{Q} \mathbf{X_1} + \lambda^{(0)} \mathbf{A}^\top \mathbf{A})^{-1} \mathbf{X_1}^\top \mathbf{Q} y, \\[1em]
\bm\eta^{(0)}(\lambda^{(0)}) = (\mathbf{X_2}^\top \mathbf{X_2})^{-1} \mathbf{X_2}^\top (y - \mathbf{X_1}\bm\beta^{(0)}(\lambda^{(0)}). 
\end{align*}

Then we can calculate the initial value $\bm\delta_{ij}^{(0)}$ using the definition of $\bm\delta_{ij}$:

\begin{align*}
\bm\delta_{ij}^{(0)}=\bm\beta_i^{(0)}-\bm\beta_j^{(0)}.
\end{align*}

Initial value of the dual variables $\bm{\alpha}$ is set to be $\bm{\alpha}^{(0)} = \bm{0}$.

\section{Evaluation Metrics}

{For clustering performance, we employ three widely used measures:}

\textbf{Rand Index (RI)} quantifies the proportion of correctly identified pairwise relationships between spatial units. It is defined as
\begin{equation*}
\text{RI} = \frac{\text{TP} + \text{TN}}{\text{TP} + \text{FP} + \text{FN} + \text{TN}},
\end{equation*}
where TP, TN, FP, and FN denote true positives, true negatives, false positives, and false negatives, respectively. RI values range from 0 to 1, with higher values indicating more accurate clustering.

\textbf{Clustering Accuracy (CA)} measures the proportion of spatial units correctly assigned to their true clusters:
\begin{equation*}
\text{CA} = \frac{\text{Number of Correctly Assigned Locations}}{\text{Total Number of Locations}}.
\end{equation*}

To account for the inherent label indeterminacy in unsupervised learning, estimated cluster labels are aligned with true labels using the Hungarian algorithm, as implemented in the \texttt{clue} R package \citep{Hornik2005clue} via the \texttt{solve\_LSAP()} function. This metric directly evaluates the model’s ability to recover the true spatial clustering structure.

\textbf{Relative Cluster Count (RCC)} assesses the extent to which the estimated number of clusters matches the true number:
\begin{equation*}
\text{RCC} = \frac{\text{Mean Estimated Number of Clusters}}{\text{True Number of Clusters}}.
\end{equation*}
An RCC value of 1 indicates perfect recovery of the true cluster count, whereas values greater than 1 indicate over-clustering and values less than 1 indicate under-clustering. This measure captures the model’s tendency to fragment or merge spatial clusters across simulation replicates.

To evaluate estimation accuracy, we compute the \textbf{bias} and \textbf{mean squared error (MSE)} of the estimated regression coefficients. The bias of $\bm{\beta}$ is defined as
\begin{equation*}
\text{Bias} = \left\| \frac{1}{R} \sum_{r=1}^{R} \hat{\bm{\beta}}_i^{(r)} - \bm{\beta}_i \right\|_2,
\end{equation*}
and the overall bias is obtained by averaging across all spatial units, providing a concise measure of directional and magnitude error.

The {MSE} quantifies overall estimation accuracy:
\begin{equation*}
\text{MSE} = \left\| \frac{1}{R} \sum_{r=1}^{R} \hat{\bm{\beta}}_i^{(r)} - \bm{\beta}_i \right\|_2^2.
\end{equation*}
Averaging MSE values across spatial units yields a comprehensive metric that reflects both the variance and squared bias of the estimator.

\section{Lattice simulation}

\begin{figure}[htbp]
    \centering
    \includegraphics[width=0.5\linewidth]{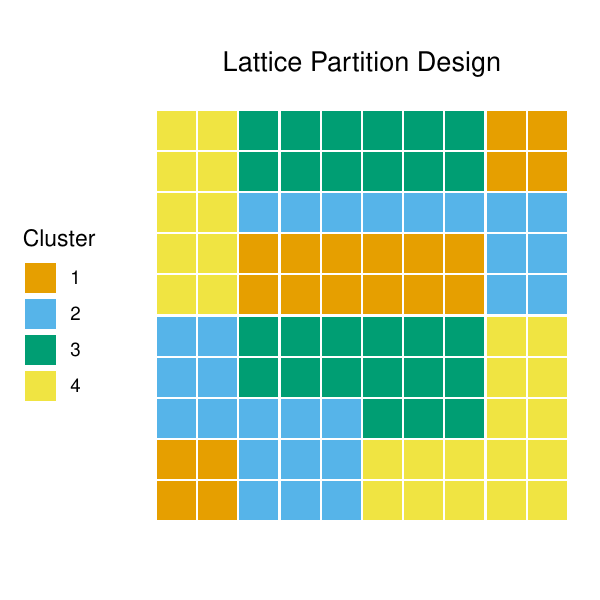}
    \caption{lattice-based partition design}
    \label{fig:latticemap}
\end{figure}

{In addition to the two simulation in main content, we include this lattice-based spatial structure to provide a controlled experimental setting in which spatial relationships are explicitly defined through regular adjacency. A key advantage of this design is that it allows the number of spatial units to be flexibly scaled, enabling systematic evaluation of model performance as the problem size increases.
The lattice configuration serves as a benchmark environment where the spatial structure is known and uniformly constructed, which facilitates fair comparison across methods. It also allows for clear control over neighborhood relationships and spatial dependence, making it possible to isolate the effects of model components without confounding from irregular geographic boundaries.
Such a structured setting is particularly useful for assessing the robustness and scalability of the proposed method, as well as for understanding how it behaves under different levels of spatial complexity.
Under this lattice setting, we evaluate how well the proposed method recovers spatially heterogeneous regimes and clustering structures under controlled conditions. In particular, we assess its ability to capture underlying spatial patterns and compare its performance with existing approaches.}

\renewcommand{\thetable}{\arabic{table}}
\begin{table}[tbp]
\centering
\caption {\label{tab:lattice_summary} Summary of lattice simulation performance} 
\vspace{0.1cm}
\resizebox{5in}{!}{%
\begin{tabular}{llrrrrrrr}
\toprule
&  & \multicolumn{3}{l}{Performance} & \multicolumn{2}{l}{Estimation-$\beta$} & \multicolumn{2}{l}{Estimation-$\eta$} \\ \cmidrule(lr){3-9}
Method       & r    & RI   & CA   & RCC   & Bias & MSE  & Bias & MSE  \\
\midrule
Constant     & -    & 0.67  & 0.49  & 0.52   & 1.52  & 3.07  & 0.64  & 0.55  \\
Pure adjacency       & -    & -     & -     & -      & -     & -     & -     & -     \\
Spatial-pairwise      & 1.5  & 0.70  & 0.53  & 0.93   & 1.03  & 1.98  & 0.47  & 0.28  \\
Spatial-adjacency & 1.5  & \textbf{0.70}  & \textbf{0.55}  & 1.25   & \textbf{0.75}  & \textbf{1.25 } & {0.17}  & {0.03}\\
\bottomrule
\end{tabular}
}
\end{table}

{As summarized in Table~\ref{tab:lattice_summary}, the proposed spatial-adjacency approach demonstrates clear advantages over competing methods, consistently yielding higher clustering precision and more accurate recovery of the underlying spatial structures. It also achieves lower estimation errors for both compositional and non-compositional coefficients, indicating superior model fit. In contrast, the pure adjacency specification failed to converge due to the extreme sparsity of its spatial weight matrix, which led to numerical instability during optimization and precluded the generation of a summarized result. The pure adjacency model defines spatial relationships solely based on whether two units share a common boundary, without considering distance or intensity of connection. This produces a binary and often extremely sparse spatial weight matrix, especially in lattice settings with limited neighborhood links. Such sparsity reduces spatial information and can lead to poorly conditioned estimation problems. As a result, model optimization becomes unstable and may fail to converge.}

%\bibliographystyle{apacite}

%\bibliography{Bib}

\end{document}